\documentclass[
    aps,
    prl,
    reprint,
    superscriptaddress,
    nofootinbib,
    floatfix
]{revtex4-2}

\usepackage[T1]{fontenc}
\usepackage[utf8]{inputenc}
\usepackage{amsmath,amssymb,mathtools}
\usepackage{bm}
\usepackage{graphicx}
\usepackage{dcolumn}
\usepackage{booktabs}
\usepackage{xcolor}
\usepackage{hyperref}
\usepackage{microtype}
\usepackage{todonotes}
\newcommand{\msbar}{\bar{\mu}_s}

\begin{document}

\title{Gravitational-Wave Constraints on $R^2$ Gravity}

\author{Pawan Kumar Gupta}
\email{pkgupta@camk.edu.pl}
\affiliation{Nicolaus Copernicus Astronomical Center, Polish Academy of Sciences, 00-716, Warsaw, Poland}

\author{Laura Sagunski}
\email{sagunski@itp.uni-frankfurt.de}
\affiliation{Institute for Theoretical Physics, Goethe University,
60438 Frankfurt am Main, Germany}

\date{\today}

\begin{abstract}
Gravitational waves from compact binary coalescences provide a unique opportunity to test general relativity in the strong-field regime. Quadratic $f(R)=R+aR^2$ gravity is among the simplest higher-curvature extensions of general relativity and introduces a massive scalar degree of freedom whose finite interaction range can modify compact-binary dynamics and gravitational-wave emission. We map this theory onto the effective-field-theory waveform for generic massive scalar-tensor gravity, obtaining a theory-specific inspiral phase correction governed by the single coupling parameter $a$. Using Bayesian analyses of the binary-neutron-star events GW170817 and GW190425,  we obtain a combined constraint on the quadratic-curvature coupling of $a\lesssim6.2\times10^6\,\mathrm{m}^2$ at $90\%$ credibility, corresponding to a lower bound on the scalar mass of $m_s\gtrsim3.2\times10^{-11}\,\mathrm{eV}$. This constitutes the strongest astrophysical constraint on the quadratic-curvature coupling, improving the previous best bound by approximately five orders of magnitude and demonstrating the sensitivity of binary-neutron-star inspirals to finite-range scalar interactions in the strong-field regime.

\end{abstract}

\maketitle

\textbf{\textit{Introduction.}} Gravitational waves (GWs) from compact-binary coalescences, observed by Advanced LIGO~\cite{LIGOScientific:2014pky}, Advanced Virgo~\cite{VIRGO:2014yos}, and KAGRA~\cite{KAGRA:2018plz}, provide a unique probe of general relativity (GR) in the strong-field, highly dynamical regime~\cite{Will:2014kxa,Berti:2015itd,Yunes:2016jcc,LIGOScientific:2016lio,LIGOScientific:2018dkp,LIGOScientific:2020tif}. Tests of GR with these observations include theory-agnostic searches for deviations in the inspiral, merger, and ringdown signals~\cite{Arun:2006hn,Yunes:2009ke,Mishra:2010tp,Li:2011cg,Shoom:2021mdj}, as systematically performed by the LIGO--Virgo--KAGRA (LVK) collaborations~\cite{LIGOScientific:2019fpa,LIGOScientific:2021sio,LIGOScientific:2021nrg,LIGOScientific:2026fcf,LIGOScientific:2026gwtc5gr}, as well as theory-specific tests that directly constrain the parameters of particular extensions of GR~\cite{Gupta:2021rod,Yagi:2011xp,Yagi:2011ze,Yagi:2012vf,Yunes:2011we,Yagi:2012ya,Silva:2017uqg,Roy:2025xih,Xie:2024xex,Sanger:2024axs}. Many extensions of GR introduce additional gravitational degrees of freedom, which can endow compact objects with scalar or vector charges and lead to observable modifications in GW signals~\cite{Berti:2018cxi,Sotiriou:2015pka,vanGemeren:2023rhh,Gupta:2022spq,Gupta:2026mqk,Sagunski:2017nzb,Huang:2018pbu,Zhang:2021mks}. 

Among the most extensively studied extensions of GR exhibiting an additional scalar degree of freedom are $f(R)$ theories of gravity~\cite{Sotiriou:2008rp,DeFelice:2010aj}. A particularly simple and theoretically well-motivated realization is quadratic $f(R)$ gravity, $f(R)=R+aR^2$, which introduces a single additional parameter $a$ beyond GR and is dynamically equivalent to a scalar-tensor theory with a massive scalar degree of freedom. Unlike scalar-tensor theories that admit spontaneous scalarization~\cite{Damour:1993hw,Damour:1992we}, $R^2$ gravity does not exhibit such a mechanism. 

The $R^2$ term arises naturally among the higher-curvature corrections to the Einstein-Hilbert action and is particularly well known from the Starobinsky model of inflation, where the additional scalar degree of freedom drives an early phase of accelerated expansion~\cite{Starobinsky:1980te}. Such curvature corrections are also generically expected in the low-energy effective-field-theory (EFT) description of gravity~\cite{Donoghue:1994dn,Burgess:2003jk}. More general $f(R)$ models have also been extensively investigated in the context of late-time cosmic acceleration~\cite{Carroll:2003wy,Hu:2007nk}.

Compact objects provide a natural strong-field regime for probing higher-curvature corrections such as $R^2$ gravity. For black holes, however, the stationary vacuum solutions coincide with those of GR under the usual assumptions and therefore do not provide a direct probe of the additional scalar degree of freedom. Neutron stars (NSs) offer a qualitatively different avenue: the additional massive scalar is sourced by matter inside the star and can develop a nontrivial configuration, modifying its structure and gravitational interaction~\cite{Cooney:2009rr,Arapoglu:2010rz,Orellana:2013gn,Staykov:2014mwa,Diedrichs:2023trk,Jockel:2023rrm}. 

These effects can become particularly relevant in binary neutron-star (BNS) systems, where the finite-range scalar interaction can modify the orbital dynamics and leave characteristic imprints on the emitted GWs~\cite{Sagunski:2017nzb}. BNS systems are especially promising for such tests because their long inspirals accumulate a large number of observable cycles in the sensitive band of ground-based detectors. Neutron-star–black-hole (NSBH) binaries provide a complementary probe, as dipolar scalar radiation introduces the leading correction to the GW phase when scalar emission is kinematically allowed.

Previous studies have investigated neutron-star structure, binary dynamics, finite-range scalar interactions, and possible GW signatures in $f(R)$ gravity~\cite{Cooney:2009rr,Arapoglu:2010rz,Orellana:2013gn,Staykov:2014mwa,Sagunski:2017nzb}. GW observations have also been used to place constraints on $f(R)$ gravity using GW170817 based on a modification of the chirp mass~\cite{Jana:2018djs}. However, a Bayesian parameter-estimation analysis employing a theory-specific inspiral waveform with explicit $R^2$-gravity phase corrections remains unexplored.

Laboratory tests of the gravitational inverse-square law yield a stringent bound, $a\lesssim10^{-10}\,{\rm m}^2$, from the E\"ot-Wash experiment ~\cite{Kapner:2006si}, while astrophysical tests such as Gravity Probe B allow values as large as $a\lesssim5\times10^{11}\,{\rm m}^2$ and the measurements of the precession of the pulsar B in the PSR J0737-3039 system provide instead the limit $a \lesssim 2.3 \times 10^{15}\,\mathrm{m}^2$~\cite{Breton:2008xy, Naf:2010zy}. The interpretation of bounds obtained in such different environments can, however, be model-dependent: screening mechanisms in more general $f(R)$ theories can suppress scalar-mediated fifth forces and thereby evade local constraints~\cite{Khoury:2003aq,Khoury:2003rn}. This motivates independent tests in the strong-field regime of compact binaries. 

In this Letter, we specialize the EFT waveform for
generic massive scalar-tensor binaries~\cite{Diedrichs:2023} to
quadratic $f(R)$ gravity. We show that the scalar charges, induced
couplings, self-interactions, and scalar mass entering the generic
waveform are all fixed by the single parameter $a$, yielding a
one-parameter extension of the GR inspiral waveform with scalar-induced phase corrections through 0 post-Newtonian (PN) order. We then perform a Bayesian analysis of the BNS events GW170817 and GW190425, obtaining a combined constraint on $a$ that improves the strongest existing astrophysical bound from Gravity Probe B by approximately five orders of magnitude.

Throughout, we use natural units $c=\hbar=1$, keeping $G$ explicit in the
analytic derivations, and adopt the metric signature $(-,+,+,+)$.
We restore $c$ and $\hbar$ explicitly in formulas used to convert
to physical masses, lengths, and frequencies. 

\textbf{\textit{Impact of finite-range scalar forces on binary mergers.}}
Before specializing to quadratic $f(R)$ gravity, we outline how
finite-range scalar forces can affect compact-binary dynamics. In theories where a massive scalar couples universally to matter, NSs can acquire appreciable scalar charges~\cite{Barausse:2012da, Sagunski:2017nzb}. The resulting Yukawa interaction has a range set by the scalar's Compton wavelength,
$\lambda_s=1/m_s$. Although the force is exponentially suppressed at separations much larger than $\lambda_s$, it can modify the orbital evolution and produce observable GW phase shifts at smaller separations. BNS inspirals therefore probe the strength and range of this interaction; at leading order, its strength depends on the product of the stars' effective scalar charges~\cite{Barausse:2012da}.

The sensitivity of binaries containing black holes depends on whether the black holes carry scalar charge. In theories where they do not, the leading interaction between scalar charges is absent in both binary-black-hole (BBH) and NSBH systems. NSBH binaries can nevertheless lose energy through scalar dipole radiation because the two objects have different scalar charge-to-mass ratios. For quasicircular orbits, this radiation can propagate only when the orbital angular frequency exceeds the scalar mass
$m_s$.
BNS observations therefore offer opportunities to constrain the strength and range of finite-range scalar forces~\cite{Sagunski:2017nzb}, while NSBH observations can provide complementary tests where scalar radiation is kinematically allowed and measurable.

\textbf{\textit{$R^2$ gravity as a one-parameter scalar-tensor theory.}}
We now specialize to quadratic $f(R)$ gravity, $f(R)=R+aR^2$, in which a single coupling parameter $a$ determines the scalar mass and interaction range. 

The Jordan-frame action of quadratic $f(R)$ gravity is
\begin{equation}
S=
\frac{1}{16\pi G}
\int d^4x\,\sqrt{-g}\,
\left(R+aR^2\right)
+S_{\rm matter}(g_{\mu\nu},\chi),
\label{eq:R2_action}
\end{equation}
where $R$ is the Ricci scalar associated with $g_{\mu\nu}$ and $\chi$
collectively denotes the matter fields. The coupling parameter must be positive, $a>0$, with $a\to0$ recovering GR. A negative $a$ would lead to tachyonic instabilities or the appearance of
ghosts. Quadratic $f(R)$ gravity is
dynamically equivalent to a massive scalar-tensor theory
~\cite{Sotiriou:2006hs,Yazadjiev:2014cza} and, in its Jordan-frame representation, to a
Brans--Dicke theory with vanishing Brans--Dicke parameter and a self-interaction potential~\cite{Wands:1993uu}. Introducing the Einstein-frame
metric
\begin{equation}
g_{\mu\nu}^*=(1+2aR)g_{\mu\nu}
\end{equation}
and the scalar field
\begin{equation}
\varphi=\frac{\sqrt{3}}{2}\ln(1+2aR),
\label{eq:conformal_coupling}
\end{equation}
the action becomes
\begin{align}
S_{\rm E}
={}&
\frac{1}{16\pi G}
\int d^4x\,\sqrt{-g_*}
\left[
R_*
-2g_*^{\mu\nu}\partial_\mu\varphi\partial_\nu\varphi
-V(\varphi)
\right]
\nonumber\\
&+
S_{\rm matter}
\left(e^{-2\varphi/\sqrt{3}}g_{\mu\nu}^{*},\chi\right),
\label{eq:einstein_action}
\end{align}
where
\begin{equation}
V(\varphi)=
\frac{1}{4a}
\left(1-e^{-2\varphi/\sqrt{3}}\right)^2.
\label{eq:einstein_potential}
\end{equation}
Expanding the potential about its minimum at $\varphi=0$ gives
the scalar mass parameter in natural units and the corresponding interaction range,
\begin{equation}
m_s^2=\frac{1}{6a},
\qquad
\lambda_s=m_s^{-1}=\sqrt{6a}.
\label{eq:scalar_mass}
\end{equation}
Thus, the scalar mass and interaction range are fixed by $a$, leaving a single additional theory parameter in the inspiral model. In the GR limit, $a\rightarrow 0$, corresponding to $m_s\rightarrow\infty$, the scalar interaction becomes short-ranged and decouples.

The scalar field modifies the binary dynamics through its coupling to matter~\cite{Yazadjiev:2014cza,Sagunski:2017nzb},
\begin{equation}
\Box_{*}\varphi=\frac{1}{4}\frac{dV}{d\varphi}
+\frac{4\pi G}{\sqrt{3}}T_{*},
\label{eq:scalar_eom}
\end{equation}
where $T_{*}=-\rho_{*}+3p_{*}$ is the matter trace in the Einstein frame. Its generally nonzero value inside NSs sources a scalar field, allowing BNS systems to probe finite-range scalar forces and, when kinematically allowed, scalar radiation. NSBH binaries can likewise emit scalar radiation despite vanishing scalar charge of the black hole. For an isolated vacuum BBH system with no pre-existing scalar field or scalar waves, $\varphi$ remains zero throughout the inspiral and merger. The binary therefore evolves as in general relativity and cannot constrain $a$ through these scalar effects.

\textbf{\textit{Gravitational waves in $R^2$ gravity from compact binaries.}}
We now construct the GW inspiral waveform for compact binaries in quadratic $f(R)$ gravity by matching its scalar-tensor representation to the EFT framework of Ref.~\cite{Diedrichs:2023}. This framework incorporates conservative dynamics through 1PN order and scalar-induced corrections to the GW phase through 0PN order. We consider the BNS system as the general case and then obtain the NSBH waveform by taking the appropriate limit for black-hole.

We begin by describing the compact objects through the point-particle matter action
\begin{equation}
S_{\rm matter}(g_{\mu\nu},\chi)
=
-\sum_{A=1,2}m_A\int d\tau_A .
\label{eq:matter_jordan}
\end{equation}
where $m_A$ is the mass of compact object $A$, and $d\tau_A^2=-g_{\mu\nu}dx_A^\mu dx_A^\nu$ denotes its proper-time interval. Under the conformal transformation
$d\tau_A=e^{-\varphi/\sqrt{3}}d\tau_{*,A}$, the matter action in the Einstein frame becomes
\begin{equation}
S_{\rm matter}
\left(e^{-2\varphi/\sqrt{3}}g_{\mu\nu}^{*},\chi\right)
=
-\sum_{A=1,2}m_A
\int e^{-\varphi/\sqrt{3}}\,d\tau_{*,A}.
\label{eq:compactmatteraction}
\end{equation}

To obtain the corresponding waveform, we employ the generic massive scalar-tensor EFT of Ref.~\cite{Diedrichs:2023}, whose action takes the form
\begin{equation}
S_{\rm EFT}=S_{\rm EH}+S_{\phi}+S_{\rm pp},
\label{eq:EFT_action}
\end{equation}
where
\begin{align}
S_{\rm EH}
&=2M_{\rm Pl}^2
\int d^4x\,\sqrt{-g_*}\,R_*,
\nonumber\\
S_{\phi}
&=\int d^4x\,\sqrt{-g_*}
\left[
-\frac12(\partial\phi)^2-V(\phi)
\right],
\label{eq:EFT_bulk}
\end{align}
and
\begin{equation}
S_{\rm pp}
=-\sum_{A=1,2}\int d\tau_A
\left[
m_A
+q_A\frac{\phi}{M_{\rm Pl}}
+p_A\left(\frac{\phi}{M_{\rm Pl}}\right)^2
\right].
\label{eq:eft_worldline_action}
\end{equation}
Here, $M_{\rm Pl}^2=1/(32\pi G)$, and $q_A$ and $p_A$ denote the linear and quadratic scalar couplings of each compact object, respectively. The scalar potential is expanded up to 1PN order, with $\phi \sim \mathcal{O}(1/c^2)$,
\begin{equation}
V(\phi)
=
\frac12m_s^2\phi^2
+\frac{M_{\rm Pl}c_3}{3!}\phi^3
+\mathcal{O}(\phi^4).
\label{eq:potential_expansion}
\end{equation}
The quartic term enters only at higher order in the scalar couplings than $c_3$ and is therefore neglected here.

To specialize the generic EFT framework to quadratic $f(R)$ gravity, we
match the Einstein-frame action in Eq.~\eqref{eq:einstein_action} and
the matter coupling in Eq.~\eqref{eq:compactmatteraction} to
Eqs.~\eqref{eq:EFT_bulk}--\eqref{eq:potential_expansion}.
Canonical normalization of the scalar kinetic term gives
\begin{equation}
\phi=2\sqrt{2}\,M_{\rm Pl}\varphi.
\label{eq:field_redefinition}
\end{equation}
In terms of the canonically normalized field, the matter action Eq.~\eqref{eq:compactmatteraction} becomes
\begin{equation}
S_{\rm matter}
=
-\sum_{A=1,2}m_A
\int d\tau_A\,
e^{-\phi/(2\sqrt{6}M_{\rm Pl})}.
\label{eq:matter_canonical}
\end{equation}
Expanding the exponential and matching the coefficients to the
point-particle action in Eq.~\eqref{eq:eft_worldline_action}, we obtain
\begin{equation}
q_A=-\frac{m_A}{2\sqrt{6}},
\qquad
p_A=\frac{m_A}{48}.
\label{eq:worldline_coefficients}
\end{equation}
These couplings satisfy $p_A=q_A^2/(2m_A)$ and yield
a universal scalar charge-to-mass ratio
$q_A/m_A=-1/(2\sqrt{6})$ within the adopted
point-particle approximation. For finite-size neutron
stars, however, the effective scalar charges can depend
on their compactness and internal structure.

The scalar self-interactions are determined by the Einstein-frame
potential Eq.~\eqref{eq:einstein_potential}, which, in terms of the canonically normalized field, reads
\begin{equation}
V(\phi)
=
\frac{M_{\rm Pl}^2}{2a}
\left(
1-e^{-\phi/(\sqrt{6}M_{\rm Pl})}
\right)^2.
\label{eq:canonical_potential}
\end{equation}
Expanding this potential around $\phi=0$ and comparing with
Eq.~\eqref{eq:potential_expansion}, we obtain
\begin{align}
m_s^2
=\frac{1}{6a},
\quad c_3
=-\sqrt{\frac32}\frac{m_s^2}{M_{\rm Pl}^2}.
\label{eq:self_couplings}
\end{align}

Equations~\eqref{eq:worldline_coefficients} and
\eqref{eq:self_couplings} establish the matching between quadratic
$f(R)$ gravity and the generic massive scalar-tensor EFT. The scalar
mass, self-interactions, and point-particle couplings are thereby
fixed by the single quadratic-curvature parameter $a$, yielding a
theory-specific, one-parameter extension of the GR inspiral waveform.

Using the matching relations derived above and inserting it into the generic massive scalar-tensor waveform~\cite{Diedrichs:2023}, we construct the
frequency-domain inspiral waveform in quadratic $f(R)$ gravity,
including scalar-induced phase corrections through 0PN
order,
\begin{equation}
\tilde h(f)
=
\mathcal{A}(f)
\exp\left[
i\Psi_{\rm GR}(f)
+i\Delta\Psi_{R^2}(f; a)
\right],
\label{eq:waveform}
\end{equation}
where $\mathcal{A}(f)$ and $\Psi_{\rm GR}(f)$ denote the GR amplitude
and phase, respectively. The additional $R^2$ phase correction, $\Delta\Psi_{R^2}(f; a)$, is given by
\begin{align}
\Delta\Psi_{R^2}(f; a)
={}&
\Psi_E
+\Theta\left(\frac{v^3}{\bar{\mu}_s}-1\right)\Psi_{\ell=1}
+\Theta\left(\frac{v^3}{\bar{\mu}_s}-\frac{1}{2}\right)\Psi_{\ell=2}.
\label{eq:ST_phase}
\end{align}

Here, $\Psi_E$ is the conservative Yukawa contribution, and $\Psi_{\ell=1}$ and $\Psi_{\ell=2}$ are the scalar dipole and quadrupole radiation contributions (see Appendix~\ref{app:R2_phase}). $\Theta$ is the Heaviside step function and $M=m_1+m_2$. In natural units, the dimensionless phase variables are
$v=(\pi GMf)^{1/3}$ and $\bar\mu_s=GM\mu_s$, where $\mu_s\equiv m_s$ is the scalar mass in natural units.\footnote{With $c$ restored, $v=(\pi GMf/c^3)^{1/3}$, $\bar\mu_s=GM\mu_s/c^2$, and $\mu_s=m_s c/\hbar$.} These coefficients follow from substituting the matching relations
Eqs.~(\ref{eq:worldline_coefficients}) and (\ref{eq:self_couplings}) into the generic EFT
waveform. 
Scalar dipole radiation first contributes at $-1$PN order, whereas $\Psi_E$ and scalar quadrupole radiation start at $0$PN order. The finite interaction range and radiation thresholds give the scalar correction a non-polynomial frequency dependence.

The phase correction given in Eq.~\eqref{eq:ST_phase} is for any generic compact binary, the specific correction for BNS systems is given in Eq.~\eqref{eq:app_complete_R2_phase_BNS}, and for NSBH systems is given in Eq.~\eqref{eq:app_complete_R2_phase_NSBH}. For BNS systems, the universal scalar charge-to-mass ratio, $q_A/m_A=-1/(2\sqrt{6})$ (within the point-particle approximation), causes the dipole contributions to vanish, leaving the conservative Yukawa interaction and scalar quadrupole radiation as the leading corrections. In contrast, for NSBH systems, the vanishing black-hole scalar charge eliminates the leading conservative Yukawa interaction, while the unequal scalar charge-to-mass ratios generate dipolar radiation.

\textbf{\textit{Bayesian analysis of GW events.}}
We consider seven compact-binary coalescences observed by the LVK Collaboration, including two BNS events, GW170817~\cite{Abbott:2017vwq} and GW190425~\cite{Abbott:2020uma}, and five NSBH events or candidates: GW200105, GW200115~\cite{Abbott:2021gcn}, GW191219~\cite{Abbott:2021vaw}, GW230518, and GW230529~\cite{Abac:2024zpt}.

\begin{figure}
    \centering
    \includegraphics[width=1.0\linewidth]{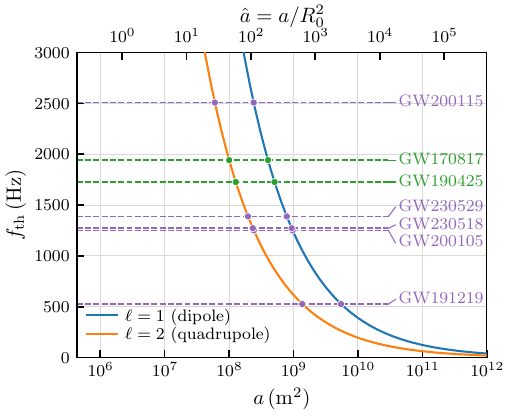}
    \caption{Scalar-radiation activation frequencies as functions of
    the quadratic-curvature coupling $a$ (bottom axis) and its dimensionless counterpart $\hat a=a/R_0^2$ (top axis).    The solid blue and orange curves represent the kinematic thresholds for scalar dipole ($\ell=1$) and quadrupole ($\ell=2$) radiation, respectively. The horizontal lines indicate the estimated merger frequencies of individual BNS (green) and NSBH (purple) events. The intersections, marked by circles, identify the minimum coupling required for the corresponding scalar-radiation channel to become accessible before merger.}
    \label{fig:scalar_activation}
\end{figure}

Since stationary black holes carry no scalar charge
under the usual assumptions, the leading scalar-mediated
conservative interaction is absent in NSBH systems, leaving
dipolar scalar radiation as the dominant correction.
However, dipolar emission becomes kinematically accessible only
when the GW frequency exceeds the threshold
$f_{\rm th}^{(1)}=c/(\pi\sqrt{6a})$. 

For the observed NSBH systems considered here, this threshold is reached before merger only for relatively large values of $a$, typically $a\gtrsim10^8$–$10^9,\mathrm{m}^2$, corresponding to $\hat a=a/R_0^2$, where $R_0=GM_\odot/c^2=1.47664,\mathrm{km}$, as illustrated in Fig.~\ref{fig:scalar_activation}. The merger frequency has been computed for each BNS event, considering nonspinning NSs from~\cite{Dietrich:2017aum}. For NSBH events we use the remnant's dominant ringdown frequency~\cite{Thompson2020,London2019},
computed from the median final mass $M_f$ and spin $\chi_f$ in the GW Transient Catalog releases GWTC-3~\cite{GWTC3} and GWTC-4.0~\cite{GWTC4}.

For NSBH events, the scalar radiative correction vanishes below the threshold and therefore carries no information about small couplings. Even for very large values of $\hat a\sim3000$, the phase correction for NSBH systems is only of the order of a few radians, as shown in Fig.~\ref{fig:phase_2D_map_NSBH}. In contrast, BNS systems have a conservative Yukawa contribution to the phase. This contribution has no radiation threshold: it is present at all separations and is only exponentially suppressed for $r\gg\lambda_s$. For BNS systems, the phase
correction is of the order of tens of radians for $\hat a\sim5$, as shown in Fig.~\ref{fig:phase_2D_map}. The BNS posteriors are also consistent with the GR limit. This motivates our choice of BNS events for the Bayesian analysis.

We analyze the GW data from the BNS events
GW170817 and GW190425 using Bayesian parameter estimation,
assuming stationary Gaussian detector noise. The parameter space includes the standard BNS source parameters and
the dimensionless quadratic-curvature coupling $\hat a=a/R_0^2$ with
$R_0=GM_\odot/c^2=1.47664\,{\rm km}$. To obtain a joint constraint on $\hat a$, we multiply the
marginalized posteriors from GW170817 and GW190425, accounting for the common prior~\cite{Zimmerman:2019wzo}.

The parameter estimation is performed with the
\textsc{Bilby} framework~\cite{Ashton:2018jfp} using the nested-sampling algorithm
\textsc{Dynesty}~\cite{Speagle:2019ivv}. We employ the
\texttt{IMRPhenomPv2\_NRTidalv2} waveform as the GR baseline,
supplemented by the $R^2$ phase correction of
Eq.(~\ref{eq:app_complete_R2_phase_BNS}). The \texttt{IMRPhenomPv2\_NRTidalv2} waveform model incorporates spin-precession effects and tidal corrections to the BNS inspiral from GR~\cite{Hannam:2013oca,Dietrich:2018uni,Dietrich:2019kaq}. The tidal corrections are expected to differ in $R^2$ gravity, because $a$ changes the stellar structure~\cite{Yazadjiev:2018xxk}. We do not include these modifications here, as they require a separate analysis that we leave for future work.
The sampled parameters include the detector-frame chirp mass,
mass ratio, component spins, tidal deformabilities, luminosity
distance, sky position, binary orientation, coalescence time and phase,
together with the $R^2$-gravity parameter $\hat a$.
We adopt the standard low-spin prior
($|\chi_{1,2}|<0.05$), the publicly released calibration envelopes,
and the detector settings used in the LVK analysis of the GW170817 and GW190425 events~\cite{LIGOScientific:2019lzm,ligo_scientific_collaboration_and_virgo_2022_6513631}. The prior on the modified-gravity parameter is taken to be
uniform over the chosen prior range
$\hat a \in [0, 100]$, with the GR limit recovered at
$\hat a=0$~(Fig.~\ref{fig:phase_2D_map}).

\begin{figure}
    \centering
    \includegraphics[width=\columnwidth]
    {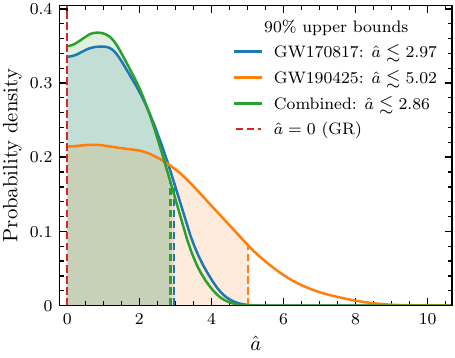}
    \caption{
    Marginalized posterior probability densities for the dimensionless
    quadratic-curvature coupling $\hat a$, obtained from Bayesian
    analyses of the BNS events GW170817 and GW190425 and their
    combination. The red dashed line indicates the GR limit,
    $\hat a=0$. The posteriors are consistent with GR and yield
    90\% credible upper bounds of $\hat a\lesssim2.97$ for GW170817,
    $\hat a\lesssim5.02$ for GW190425, and
    $\hat a\lesssim2.86$ for the combined analysis.
    }
    \label{fig:ahat_posterior}
\end{figure}
\textbf{\textit{Results.}}
The marginalized posterior distributions for the dimensionless
quadratic-curvature coupling $\hat a$, obtained from Bayesian analyses
of the BNS events GW170817 and GW190425, are shown in
Fig.~\ref{fig:ahat_posterior}. Both posteriors are consistent with
the GR limit, $\hat a=0$. We obtain 90\% credible upper bounds of
$\hat a\lesssim2.97$ for GW170817 and $\hat a\lesssim5.02$ for
GW190425. Combining the two events yields a tighter constraint of
$\hat a\lesssim2.86$.

These bounds correspond, through $a=\hat a R_0^2$, to
$a\lesssim6.5\times10^6\,{\rm m}^2$ for GW170817 and
$a\lesssim9.8\times10^6\,{\rm m}^2$ for GW190425.
The tighter constraint from GW170817 is consistent with its higher
network signal-to-noise ratio (SNR) of 32.4~\cite{Abbott:2017vwq}, compared
with 12.9 for GW190425~\cite{Abbott:2020uma}. The combined analysis gives
\begin{equation}
    a\lesssim6.24\times10^6\,{\rm m}^2,
\end{equation}
improving the previous
best astrophysical bound from Gravity Probe B,
$a\lesssim5\times10^{11}\,{\rm m}^2$, by a factor of
approximately $8.0\times10^4$, or nearly five orders of magnitude.

The constraints on $a$ can equivalently be expressed as lower bounds
on the scalar mass and upper bounds on its Compton wavelength.
For GW170817, we obtain
$m_s\gtrsim3.0\times10^{-11}\,{\rm eV}$ and
$\lambda_s\lesssim6.6\,{\rm km}$, while GW190425 gives
$m_s\gtrsim2.5\times10^{-11}\,{\rm eV}$ and
$\lambda_s\lesssim8.1\,{\rm km}$.
The combined analysis yields
$m_s\gtrsim3.2\times10^{-11}\,{\rm eV}$ and
$\lambda_s\lesssim6.1\,{\rm km}$.
These constraints probe a scalar-mass regime higher than those
typically considered in other scalar-field searches
~\cite{Roy:2025qaa,Xie:2024xex}.
The corresponding Compton wavelengths are comparable to neutron-star
radii, highlighting the sensitivity of BNS inspirals to
short-range scalar interactions.

\textbf{\textit{Discussion.}}
We have mapped quadratic $f(R)=R+aR^2$ gravity onto the
EFT framework of generic massive scalar-tensor
gravity, obtaining the scalar charges, induced couplings, and
self-interaction coefficients in terms of the single parameter $a$.
Using these matching relations, we construct the GW
inspiral waveform, incorporating scalar-induced corrections from
both conservative dynamics and radiation through 0PN order.
In this theory, stationary black holes have GR vacuum solutions and carry no scalar charge, whereas NSs can couple to
the scalar field, making BNS and NSBH systems promising probes
of quadratic-curvature corrections.

For BNS systems, the leading scalar-induced phase corrections arise from the conservative Yukawa interaction and scalar quadrupole radiation, while for NSBH systems, the vanishing black-hole scalar charge eliminates the leading conservative
Yukawa interaction, leaving scalar dipole and quadrupole
radiation as the dominant corrections. Unlike the conservative
interaction, scalar radiation requires the GW frequency
to exceed the corresponding kinematic threshold.
To assess the implications of these thresholds, we examine
the characteristic merger frequencies of seven GW events, including two BNS and five NSBH systems. We find that scalar
radiation in NSBH systems becomes relevant only for relatively
large couplings, $a\sim10^8\,{\rm m}^2$, whereas the
conservative Yukawa correction in BNS systems has no such
frequency threshold. This makes BNS inspirals particularly
promising for constraining $a$.

\begin{figure}
    \centering
    \includegraphics[width=\linewidth]{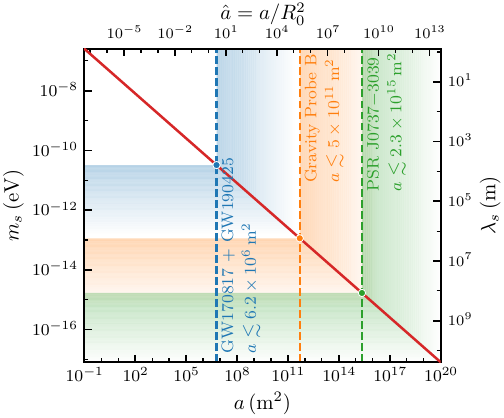}
    \caption{Comparison of the combined constraint on the quadratic-curvature
    coupling $a$ from GW170817 and GW190425 with existing
    astrophysical bounds.
    }
    \label{fig:R2_mass_vs_a_exclusion}
\end{figure}

We therefore perform Bayesian analyses of the BNS events
GW170817 and GW190425. The inferred posteriors are consistent
with GR, and their combination yields a 90\% credible upper
bound of $a\lesssim6.24\times10^6\,{\rm m}^2$. As shown in Fig.~\ref{fig:R2_mass_vs_a_exclusion}, this improves
the previous astrophysical constraints from Gravity Probe B
and PSR J0737--3039 by approximately five and eight orders
of magnitude, respectively. The improvement reflects the
sensitivity of the accumulated inspiral phase to finite-range
scalar interactions at separations much smaller than those
probed by earlier astrophysical tests.

Laboratory experiments provide complementary constraints,
including the Eöt-Wash bound
$a\lesssim10^{-10}\,{\rm m}^2$~\cite{Naf:2010zy}.
These upper limits probe different length scales and
observables and are mutually consistent. Although
environmental screening can modify the comparison in more
general $f(R)$ theories~\cite{Khoury:2003aq}, its relevance
must be established for the specific theory under
consideration. Our results therefore complement laboratory
and space-based measurements by probing quadratic-curvature
corrections through the dynamics of strongly self-gravitating
compact binaries.

Future ground-based detectors, including the Einstein Telescope~\cite{Punturo_2010,Hild:2010id} and Cosmic
Explorer~\cite{LIGOScientific:2016wof,Reitze:2019iox}, will provide improved sensitivity and access to lower
frequencies, allowing more inspiral cycles to be observed at higher SNRs and potentially strengthening constraints on $a$. More importantly, these observations may help resolve the degeneracy between gravity and the neutron-star equation of state in $R^2$ gravity~\cite{ET:2019dnz,ET:2025xjr}. In GR, the leading adiabatic tidal contribution enters at 5PN order~\cite{Flanagan:2007ix,Hinderer:2016eia}, while in $R^2$ gravity, the coupling $a$ modifies the neutron-star tidal Love numbers~\cite{Yazadjiev:2018xxk}. Beyond adiabatic tides, resonant tidal excitations are also important for third-generation detectors in extracting information about the equation of state ~\cite{Steinhoff:2016rfi,Gupta:2023oyy,Gupta:2022qgg,Gupta:2020lnv}. This motivates the development of more accurate waveform models incorporating higher-PN $R^2$ phase corrections, together with both adiabatic and resonant tidal effects in $R^2$ gravity.

\begin{acknowledgments}
We thank Micha\l{} Bejger, Badri Krishan, and Tanja Hinderer for useful discussions and comments on the draft. P.K.G.\ is partially supported by the Polish National Science Centre (NCN) 
through Grant No.~2021/43/B/ST9/01714. L.S.\ acknowledges support from
the Deutsche Forschungsgemeinschaft (DFG, German Research Foundation) through
the CRC-TR 211 `Strong-interaction matter under extreme conditions' --
project no.\ 315477589 -- TRR 211. ChatGPT was used for editorial assistance and language polishing.
\end{acknowledgments}

\bibliography{biblio, main_auto}

@article{Yazadjiev:2014cza,
    author = "Yazadjiev, Stoytcho S. and Doneva, Daniela D. and Kokkotas, Kostas D. and Staykov, Kalin V.",
    title = "{Non-perturbative and self-consistent models of neutron stars in R-squared gravity}",
    eprint = "1402.4469",
    archivePrefix = "arXiv",
    primaryClass = "gr-qc",
    doi = "10.1088/1475-7516/2014/06/003",
    journal = "JCAP",
    volume = "06",
    pages = "003",
    year = "2014"
}

@article{Hannam:2013oca,
    author = "Hannam, Mark and Schmidt, Patricia and Bohé, Alejandro and Haegel, Leïla and Husa, Sascha and Ohme, Frank and Pratten, Geraint and Pürrer, Michael",
    title = "{A simple model of complete precessing black-hole-binary gravitational waveforms}",
    eprint = "1308.3271",
    archivePrefix = "arXiv",
    primaryClass = "gr-qc",
    doi = "10.1103/PhysRevLett.113.151101",
    journal = "Phys. Rev. Lett.",
    volume = "113",
    pages = "151101",
    year = "2014"
}

@article{Dietrich:2018uni,
    author = "Dietrich, Tim and Khan, Sebastian and Dudi, Reetika and Tichy, Wolfgang and Pannarale, Francesco and Ohme, Frank and Bernuzzi, Sebastiano and Nagar, Alessandro",
    title = "{Matter imprints in waveform models for neutron star binaries: Tidal and self-spin effects}",
    eprint = "1804.02235",
    archivePrefix = "arXiv",
    primaryClass = "gr-qc",
    doi = "10.1103/PhysRevD.99.024029",
    journal = "Phys. Rev. D",
    volume = "99",
    number = "2",
    pages = "024029",
    year = "2019"
}

@article{Dietrich:2019kaq,
    author = "Dietrich, Tim and others",
    title = "{Improving the NRTidal model for binary neutron star systems}",
    eprint = "1905.06011",
    archivePrefix = "arXiv",
    primaryClass = "gr-qc",
    doi = "10.1103/PhysRevD.100.044003",
    journal = "Phys. Rev. D",
    volume = "100",
    number = "4",
    pages = "044003",
    year = "2019"
}

@article{Xie:2024xex,
    author = "Xie, Yiqi and Chung, Adrian Ka-Wai and Sotiriou, Thomas P. and Yunes, Nicol{\'a}s",
    title = "{Bayesian Search of Massive Scalar Fields from LIGO-Virgo-KAGRA Binaries}",
    eprint = "2410.14801",
    archivePrefix = "arXiv",
    primaryClass = "gr-qc",
    doi = "10.1103/PhysRevLett.134.191402",
    journal = "Phys. Rev. Lett.",
    volume = "134",
    number = "19",
    pages = "191402",
    year = "2025"
}

@article{Yazadjiev:2018xxk,
    author = "Yazadjiev, Stoytcho S. and Doneva, Daniela D. and Kokkotas, Kostas D.",
    title = "{Tidal Love numbers of neutron stars in $f(R)$ gravity}",
    eprint = "1803.09534",
    archivePrefix = "arXiv",
    primaryClass = "gr-qc",
    doi = "10.1140/epjc/s10052-018-6285-z",
    journal = "Eur. Phys. J. C",
    volume = "78",
    number = "10",
    pages = "818",
    year = "2018"
}

@article{Barausse:2012da,
    author = "Barausse, Enrico and Palenzuela, Carlos and Ponce, Marcelo and Lehner, Luis",
    title = "{Neutron-star mergers in scalar-tensor theories of gravity}",
    eprint = "1212.5053",
    archivePrefix = "arXiv",
    primaryClass = "gr-qc",
    doi = "10.1103/PhysRevD.87.081506",
    journal = "Phys. Rev. D",
    volume = "87",
    pages = "081506",
    year = "2013"
}

@article{LIGOScientific:2018dkp,
    author = "Abbott, B. P. and others",
    collaboration = "LIGO Scientific, Virgo",
    title = "{Tests of General Relativity with GW170817}",
    eprint = "1811.00364",
    archivePrefix = "arXiv",
    primaryClass = "gr-qc",
    reportNumber = "LIGO-P1800059",
    doi = "10.1103/PhysRevLett.123.011102",
    journal = "Phys. Rev. Lett.",
    volume = "123",
    number = "1",
    pages = "011102",
    year = "2019"
}

@article{Sagunski:2017nzb,
    author = "Sagunski, Laura and Zhang, Jun and Johnson, Matthew C. and Lehner, Luis and Sakellariadou, Mairi and Liebling, Steven L. and Palenzuela, Carlos and Neilsen, David",
    title = "{Neutron star mergers as a probe of modifications of general relativity with finite-range scalar forces}",
    eprint = "1709.06634",
    archivePrefix = "arXiv",
    primaryClass = "gr-qc",
    doi = "10.1103/PhysRevD.97.064016",
    journal = "Phys. Rev. D",
    volume = "97",
    number = "6",
    pages = "064016",
    year = "2018"
}

@article{Huang:2018pbu,
    author = "Huang, Junwu and Johnson, Matthew C. and Sagunski, Laura and Sakellariadou, Mairi and Zhang, Jun",
    title = "{Prospects for axion searches with Advanced LIGO through binary mergers}",
    eprint = "1807.02133",
    archivePrefix = "arXiv",
    primaryClass = "hep-ph",
    doi = "10.1103/PhysRevD.99.063013",
    journal = "Phys. Rev. D",
    volume = "99",
    number = "6",
    pages = "063013",
    year = "2019"
}

@article{Zhang:2021mks,
    author = "Zhang, Jun and Lyu, Zhenwei and Huang, Junwu and Johnson, Matthew C. and Sagunski, Laura and Sakellariadou, Mairi and Yang, Huan",
    title = "{First Constraints on Nuclear Coupling of Axionlike Particles from the Binary Neutron Star Gravitational Wave Event GW170817}",
    eprint = "2105.13963",
    archivePrefix = "arXiv",
    primaryClass = "hep-ph",
    reportNumber = "Imperial/TP/2021/JZ/01, KCL-PH-TH-2021-25, CERN-TH-2021-061,
  LIGO-P2100161, CERN-TH-2021-061, LIGO-P2100161",
    doi = "10.1103/PhysRevLett.127.161101",
    journal = "Phys. Rev. Lett.",
    volume = "127",
    number = "16",
    pages = "161101",
    year = "2021"
}

@article{Diedrichs:2023trk,
    author = {Diedrichs, Robin Fynn and Becker, Niklas and Jockel, C\'edric and Christian, Jan-Erik and Sagunski, Laura and Schaffner-Bielich, J\"urgen},
    title = "{Tidal deformability of fermion-boson stars: Neutron stars admixed with ultralight dark matter}",
    eprint = "2303.04089",
    archivePrefix = "arXiv",
    primaryClass = "gr-qc",
    doi = "10.1103/PhysRevD.108.064009",
    journal = "Phys. Rev. D",
    volume = "108",
    number = "6",
    pages = "064009",
    year = "2023"
}

@article{Sotiriou:2008rp,
    author = "Sotiriou, Thomas P. and Faraoni, Valerio",
    title = "{f(R) Theories Of Gravity}",
    eprint = "0805.1726",
    archivePrefix = "arXiv",
    primaryClass = "gr-qc",
    doi = "10.1103/RevModPhys.82.451",
    journal = "Rev. Mod. Phys.",
    volume = "82",
    pages = "451--497",
    year = "2010"
}

@article{DeFelice:2010aj,
    author = "De Felice, Antonio and Tsujikawa, Shinji",
    title = "{f(R) theories}",
    eprint = "1002.4928",
    archivePrefix = "arXiv",
    primaryClass = "gr-qc",
    doi = "10.12942/lrr-2010-3",
    journal = "Living Rev. Rel.",
    volume = "13",
    pages = "3",
    year = "2010"
}

@article{Staykov:2014mwa,
    author = "Staykov, Kalin V. and Doneva, Daniela D. and Yazadjiev, Stoytcho S. and Kokkotas, Kostas D.",
    title = "{Slowly rotating neutron and strange stars in $R^2$ gravity}",
    eprint = "1407.2180",
    archivePrefix = "arXiv",
    primaryClass = "gr-qc",
    doi = "10.1088/1475-7516/2014/10/006",
    journal = "JCAP",
    volume = "10",
    pages = "006",
    year = "2014"
}

@article{Jockel:2023rrm,
    author = "Jockel, C{\'e}dric and Sagunski, Laura",
    title = "{Fermion Proca Stars: Vector-Dark-Matter-Admixed Neutron Stars}",
    eprint = "2310.17291",
    archivePrefix = "arXiv",
    primaryClass = "gr-qc",
    doi = "10.3390/particles7010004",
    journal = "Particles",
    volume = "7",
    number = "1",
    pages = "52--79",
    year = "2024"
}

@article{LIGOScientific:2016wof,
    author = "Abbott, Benjamin P and others",
    collaboration = "LIGO Scientific",
    title = "{Exploring the Sensitivity of Next Generation Gravitational Wave Detectors}",
    eprint = "1607.08697",
    archivePrefix = "arXiv",
    primaryClass = "astro-ph.IM",
    reportNumber = "LIGO-P1600143",
    doi = "10.1088/1361-6382/aa51f4",
    journal = "Class. Quant. Grav.",
    volume = "34",
    number = "4",
    pages = "044001",
    year = "2017"
}

@article{Hild:2010id,
    author = "Hild, S. and others",
    title = "{Sensitivity Studies for Third-Generation Gravitational Wave Observatories}",
    eprint = "1012.0908",
    archivePrefix = "arXiv",
    primaryClass = "gr-qc",
    doi = "10.1088/0264-9381/28/9/094013",
    journal = "Class. Quant. Grav.",
    volume = "28",
    pages = "094013",
    year = "2011"
}

@article{Punturo_2010,
doi = {10.1088/0264-9381/27/8/084007},
url = {https://doi.org/10.1088/0264-9381/27/8/084007},
year = {2010},
month = {apr},
publisher = {},
volume = {27},
number = {8},
pages = {084007},
author = {Punturo, M. and Abernathy, M. and others},
title = {The third generation of gravitational wave observatories and their science reach},
journal = {Classical and Quantum Gravity}
}

@article{ET:2019dnz,
    author = "Maggiore, Michele and others",
    collaboration = "ET",
    title = "{Science Case for the Einstein Telescope}",
    eprint = "1912.02622",
    archivePrefix = "arXiv",
    primaryClass = "astro-ph.CO",
    doi = "10.1088/1475-7516/2020/03/050",
    journal = "JCAP",
    volume = "03",
    pages = "050",
    year = "2020"
}

@article{Reitze:2019iox,
    author = "Reitze, David and others",
    title = "{Cosmic Explorer: The U.S. Contribution to Gravitational-Wave Astronomy beyond LIGO}",
    eprint = "1907.04833",
    archivePrefix = "arXiv",
    primaryClass = "astro-ph.IM",
    reportNumber = "LIGO-P1900316",
    journal = "Bull. Am. Astron. Soc.",
    volume = "51",
    number = "7",
    pages = "035",
    year = "2019"
}

@article{ET:2025xjr,
    author = "Abac, Adrian and others",
    collaboration = "ET",
    title = "{The Science of the Einstein Telescope}",
    eprint = "2503.12263",
    archivePrefix = "arXiv",
    primaryClass = "gr-qc",
    reportNumber = "ET-0036C-25",
    doi = "10.1088/1475-7516/2026/03/081",
    journal = "JCAP",
    volume = "03",
    pages = "081",
    year = "2026"
}

@article{Flanagan:2007ix,
    author = "Flanagan, Eanna E. and Hinderer, Tanja",
    title = "{Constraining neutron-star tidal Love numbers with gravitational-wave detectors}",
    eprint = "0709.1915",
    archivePrefix = "arXiv",
    primaryClass = "astro-ph",
    journal = "Phys. Rev. D",
    volume = "77",
    pages = "021502",
    year = "2008",
    doi = "10.1103/PhysRevD.77.021502"
}

@article{Hinderer:2016eia,
    author = "Hinderer, Tanja and others",
    title = "{Effects of neutron-star dynamic tides on gravitational waveforms within the effective-one-body approach}",
    eprint = "1602.00599",
    archivePrefix = "arXiv",
    primaryClass = "gr-qc",
    journal = "Phys. Rev. Lett.",
    volume = "116",
    pages = "181101",
    year = "2016",
    doi = "10.1103/PhysRevLett.116.181101"
}

@article{Steinhoff:2016rfi,
    author = "Steinhoff, Jan and Hinderer, Tanja and Buonanno, Alessandra and Taracchini, Andrea",
    title = "{Dynamical Tides in General Relativity: Effective Action and Effective-One-Body Hamiltonian}",
    eprint = "1608.01907",
    archivePrefix = "arXiv",
    primaryClass = "gr-qc",
    journal = "Phys. Rev. D",
    volume = "94",
    pages = "104028",
    year = "2016",
    doi = "10.1103/PhysRevD.94.104028"
}

@article{Gupta:2023oyy,
    author = "Gupta, Pawan Kumar and Steinhoff, Jan and Hinderer, Tanja",
    title = "{Effect of dynamical gravitomagnetic tides on measurability of tidal parameters for binary neutron stars using gravitational waves}",
    eprint = "2302.11274",
    archivePrefix = "arXiv",
    primaryClass = "gr-qc",
    doi = "10.1103/PhysRevD.108.124040",
    journal = "Phys. Rev. D",
    volume = "108",
    number = "12",
    pages = "124040",
    year = "2023"
}

@misc{Gupta:2022qgg,
    author = "Gupta, Pawan Kumar and Puecher, Anna and Pang, Peter T. H. and Janquart, Justin and Koekoek, Gideon and Broeck Van Den, Chris",
    title = "{Determining the equation of state of neutron stars with Einstein Telescope using tidal effects and r-mode excitations from a population of binary inspirals}",
    eprint = "2205.01182",
    archivePrefix = "arXiv",
    primaryClass = "gr-qc",
    month = "5",
    year = "2022"
}

@article{Gupta:2020lnv,
    author = "Gupta, Pawan Kumar and Steinhoff, Jan and Hinderer, Tanja",
    title = "{Relativistic effective action of dynamical gravitomagnetic tides for slowly rotating neutron stars}",
    eprint = "2011.03508",
    archivePrefix = "arXiv",
    primaryClass = "gr-qc",
    doi = "10.1103/PhysRevResearch.3.013147",
    journal = "Phys. Rev. Res.",
    volume = "3",
    number = "1",
    pages = "013147",
    year = "2021"
}

@article{Zimmerman:2019wzo,
    author = "Zimmerman, Aaron and Haster, Carl-Johan and Chatziioannou, Katerina",
    title = "{On combining information from multiple gravitational wave sources}",
    eprint = "1903.11008",
    archivePrefix = "arXiv",
    primaryClass = "astro-ph.IM",
    doi = "10.1103/PhysRevD.99.124044",
    journal = "Phys. Rev. D",
    volume = "99",
    number = "12",
    pages = "124044",
    year = "2019"
}

@article{LIGOScientific:2019lzm,
    author = "Abbott, Rich and others",
    collaboration = "LIGO Scientific, Virgo",
    title = "{Open data from the first and second observing runs of Advanced LIGO and Advanced Virgo}",
    eprint = "1912.11716",
    archivePrefix = "arXiv",
    primaryClass = "gr-qc",
    reportNumber = "LIGO-P1900206",
    doi = "10.1016/j.softx.2021.100658",
    journal = "SoftwareX",
    volume = "13",
    pages = "100658",
    year = "2021"
}

@dataset{ligo_scientific_collaboration_and_virgo_2022_6513631,
  author       = {LIGO Scientific Collaboration and  Virgo Collaboration},
  title        = {GWTC-2.1: Deep Extended Catalog of Compact Binary
                   Coalescences Observed by LIGO and Virgo During the
                   First Half of the Third Observing Run - Parameter
                   Estimation Data Release
                  },
  month        = may,
  year         = 2022,
  publisher    = {Zenodo},
  version      = {v2},
  doi          = {10.5281/zenodo.6513631},
  url          = {https://doi.org/10.5281/zenodo.6513631},
}

@article{Roy:2025qaa,
    author = "Roy, Soumen and Vicente, Rodrigo and Aurrekoetxea, Josu C. and Clough, Katy and Ferreira, Pedro G.",
    title = "{Scalar Fields around Black Hole Binaries in LIGO-Virgo-KAGRA}",
    eprint = "2510.17967",
    archivePrefix = "arXiv",
    primaryClass = "gr-qc",
    reportNumber = "MIT-CTP/5937",
    doi = "10.1103/fv9z-zkxx",
    journal = "Phys. Rev. Lett.",
    volume = "136",
    number = "19",
    pages = "191402",
    year = "2026"
}

@article{Thompson2020,
  author  = {Thompson, Jonathan E. and Fauchon-Jones, Edward and Khan, Sebastian and
             Nitoglia, Elisa and Pannarale, Francesco and Dietrich, Tim and Hannam, Mark},
  title   = {Modeling the gravitational wave signature of neutron star black hole
             coalescences: {PhenomNSBH}},
  journal = {Phys. Rev. D},
  volume  = {101},
  pages   = {124059},
  year    = {2020},
  eprint  = {2002.08383},
  archivePrefix = {arXiv}
}

@article{London2019,
  author  = {London, Lionel and Fauchon-Jones, Edward},
  title   = {On modeling for {Kerr} black holes: Basis learning, {QNM} frequencies,
             and spherical-spheroidal mixing coefficients},
  journal = {Class. Quant. Grav.},
  volume  = {36},
  pages   = {235015},
  year    = {2019},
  eprint  = {1810.03550},
  archivePrefix = {arXiv}
}

@article{GWTC3,
  author  = {{LIGO Scientific, Virgo, and KAGRA Collaborations}},
  title   = {{GWTC-3}: Compact Binary Coalescences Observed by {LIGO} and {Virgo}
             during the Second Part of the Third Observing Run},
  journal = {Phys. Rev. X},
  volume  = {13},
  pages   = {041039},
  year    = {2023},
  eprint  = {2111.03606},
  archivePrefix = {arXiv}
}

@article{GWTC4,
  author  = {{LIGO Scientific, Virgo, and KAGRA Collaborations}},
  title   = {{GWTC-4.0}: Updating the Gravitational-Wave Transient Catalog with
             Observations from the First Part of the Fourth {LIGO-Virgo-KAGRA}
             Observing Run},
  journal = {Astrophys. J. Lett.},
  volume  = {1004},
  pages   = {L22},
  year    = {2026},
  eprint  = {2508.18082},
  archivePrefix = {arXiv}
}

@article{Dietrich:2017aum,
    author = "Dietrich, Tim and Bernuzzi, Sebastiano and Tichy, Wolfgang",
    title = "{Closed-form tidal approximants for binary neutron star gravitational waveforms constructed from high-resolution numerical relativity simulations}",
    eprint = "1706.02969",
    archivePrefix = "arXiv",
    primaryClass = "gr-qc",
    doi = "10.1103/PhysRevD.96.121501",
    journal = "Phys. Rev. D",
    volume = "96",
    number = "12",
    pages = "121501",
    year = "2017"
}

@article{Speagle:2019ivv,
    author = "Speagle, Joshua S.",
    title = "{dynesty: a dynamic nested sampling package for estimating Bayesian posteriors and evidences}",
    eprint = "1904.02180",
    archivePrefix = "arXiv",
    primaryClass = "astro-ph.IM",
    doi = "10.1093/mnras/staa278",
    journal = "Mon. Not. Roy. Astron. Soc.",
    volume = "493",
    number = "3",
    pages = "3132--3158",
    year = "2020"
}

@article{Ashton:2018jfp,
    author = "Ashton, Gregory and others",
    title = "{BILBY: A User-friendly Bayesian Inference Library for Gravitational-wave Astronomy}",
    eprint = "1811.02042",
    archivePrefix = "arXiv",
    primaryClass = "astro-ph.IM",
    doi = "10.3847/1538-4365/ab06fc",
    journal = "Astrophys. J. Suppl.",
    volume = "241",
    number = "2",
    pages = "27",
    year = "2019"
}

@article{Abbott:2017vwq,
    author = "Abbott, B. P. and others",
    collaboration = "LIGO Scientific and Virgo",
    title = "{GW170817: Observation of Gravitational Waves from a Binary Neutron Star Inspiral}",
    journal = "Phys. Rev. Lett.",
    volume = "119",
    pages = "161101",
    year = "2017",
    doi = "10.1103/PhysRevLett.119.161101",
    eprint = "1710.05832",
    archivePrefix = "arXiv",
    primaryClass = "gr-qc"
}

@article{Abbott:2020uma,
    author = "Abbott, B. P. and others",
    collaboration = "LIGO Scientific and Virgo",
    title = "{GW190425: Observation of a Compact Binary Coalescence with Total Mass $\sim 3.4 M_{\odot}$}",
    journal = "Astrophys. J. Lett.",
    volume = "892",
    number = "1",
    pages = "L3",
    year = "2020",
    doi = "10.3847/2041-8213/ab75f5",
    eprint = "2001.01761",
    archivePrefix = "arXiv",
    primaryClass = "astro-ph.HE"
}

@article{Abbott:2021gcn,
    author = "Abbott, R. and others",
    collaboration = "LIGO Scientific, Virgo and KAGRA",
    title = "{Observation of Gravitational Waves from Two Neutron Star--Black Hole Coalescences}",
    journal = "Astrophys. J. Lett.",
    volume = "915",
    number = "1",
    pages = "L5",
    year = "2021",
    doi = "10.3847/2041-8213/ac082e",
    eprint = "2106.15163",
    archivePrefix = "arXiv",
    primaryClass = "astro-ph.HE"
}

@article{Abbott:2021vaw,
    author = "Abbott, R. and others",
    collaboration = "LIGO Scientific, Virgo and KAGRA",
    title = "{GWTC-3: Compact Binary Coalescences Observed by LIGO and Virgo during the Second Part of the Third Observing Run}",
    journal = "Phys. Rev. X",
    volume = "13",
    number = "4",
    pages = "041039",
    year = "2023",
    doi = "10.1103/PhysRevX.13.041039",
    eprint = "2111.03606",
    archivePrefix = "arXiv",
    primaryClass = "gr-qc"
}

@article{Abac:2024zpt,
    author = "Abac, A. G. and others",
    collaboration = "LIGO Scientific, Virgo and KAGRA",
    title = "{Observation of Gravitational Waves from the Coalescence of a 2.5--4.5 Solar Mass Compact Object and a Neutron Star}",
    journal = "Astrophys. J. Lett.",
    volume = "970",
    number = "2",
    pages = "L34",
    year = "2024",
    doi = "10.3847/2041-8213/ad5beb",
    eprint = "2404.04248",
    archivePrefix = "arXiv",
    primaryClass = "astro-ph.HE"
}

@article{Wands:1993uu,
    author = "Wands, David",
    title = "{Extended gravity theories and the Einstein-Hilbert action}",
    eprint = "gr-qc/9307034",
    archivePrefix = "arXiv",
    reportNumber = "SUSSEX-AST-93-7-1",
    doi = "10.1088/0264-9381/11/1/025",
    journal = "Class. Quant. Grav.",
    volume = "11",
    pages = "269--280",
    year = "1994"
}

@article{Breton:2008xy,
    author = "Breton, Rene P. and Kaspi, Victoria M. and Kramer, Michael and McLaughlin, Maura A. and Lyutikov, Maxim and Ransom, Scott M. and Stairs, Ingrid H. and Ferdman, Robert D. and Camilo, Fernando and Possenti, Andrea",
    title = "{Relativistic Spin Precession in the Double Pulsar}",
    eprint = "0807.2644",
    archivePrefix = "arXiv",
    primaryClass = "astro-ph",
    doi = "10.1126/science.1159295",
    journal = "Science",
    volume = "321",
    pages = "104--107",
    year = "2008"
}

@article{Kapner:2006si,
    author = "Kapner, D. J. and Cook, T. S. and Adelberger, E. G. and Gundlach, J. H. and Heckel, Blayne R. and Hoyle, C. D. and Swanson, H. E.",
    title = "{Tests of the gravitational inverse-square law below the dark-energy length scale}",
    eprint = "hep-ph/0611184",
    archivePrefix = "arXiv",
    doi = "10.1103/PhysRevLett.98.021101",
    journal = "Phys. Rev. Lett.",
    volume = "98",
    pages = "021101",
    year = "2007"
}

@article{Sotiriou:2006hs,
    author = "Sotiriou, Thomas P.",
    title = "{f(R) gravity and scalar-tensor theory}",
    eprint = "gr-qc/0604028",
    archivePrefix = "arXiv",
    doi = "10.1088/0264-9381/23/17/003",
    journal = "Class. Quant. Grav.",
    volume = "23",
    pages = "5117--5128",
    year = "2006"
}

@article{Sanger:2024axs,
    author = {S{\"a}nger, Elise M. and others},
    title = "{Tests of general relativity with GW230529: A neutron star merging with a lower mass-gap compact object}",
    eprint = "2406.03568",
    archivePrefix = "arXiv",
    primaryClass = "gr-qc",
    reportNumber = "LIGO-P2400200",
    doi = "10.1103/r43k-51yq",
    journal = "Phys. Rev. D",
    volume = "113",
    number = "8",
    pages = "084070",
    year = "2026"
}

@article{Diedrichs:2023,
    author = "Diedrichs, Robin Fynn and Schmitt, Daniel and Sagunski, Laura",
    title = "{Binary systems in massive scalar-tensor theories: Next-to-leading order gravitational wave phase from effective field theory}",
    eprint = "2311.04274",
    archivePrefix = "arXiv",
    primaryClass = "gr-qc",
    doi = "10.1103/PhysRevD.110.104073",
    journal = "Phys. Rev. D",
    volume = "110",
    number = "10",
    pages = "104073",
    year = "2024"
}

@article{Jana:2018djs,
    author        = {Jana, Soumya and Mohanty, Subhendra},
    title         = {Constraints on $f(R)$ theories of gravity from GW170817},
    eprint        = {1807.04060},
    archivePrefix = {arXiv},
    primaryClass  = {gr-qc},
    doi           = {10.1103/PhysRevD.99.044056},
    journal       = {Phys. Rev. D},
    volume        = {99},
    number        = {4},
    pages         = {044056},
    year          = {2019}
}

@article{Naf:2010zy,
    author        = {N{\"a}f, Joachim and Jetzer, Philippe},
    title         = {On the 1/c Expansion of $f(R)$ Gravity},
    eprint        = {1004.2014},
    archivePrefix = {arXiv},
    primaryClass  = {gr-qc},
    journal       = {Phys. Rev. D},
    volume        = {81},
    pages         = {104003},
    year          = {2010},
    doi           = {10.1103/PhysRevD.81.104003}
}

@article{Khoury:2003aq,
    author        = {Khoury, Justin and Weltman, Amanda},
    title         = {Chameleon Fields: Awaiting Surprises for Tests of
                     Gravity in Space},
    eprint        = {astro-ph/0309300},
    archivePrefix = {arXiv},
    journal       = {Phys. Rev. Lett.},
    volume        = {93},
    pages         = {171104},
    year          = {2004},
    doi           = {10.1103/PhysRevLett.93.171104}
}

@article{Khoury:2003rn,
    author        = {Khoury, Justin and Weltman, Amanda},
    title         = {Chameleon Cosmology},
    eprint        = {astro-ph/0309411},
    archivePrefix = {arXiv},
    journal       = {Phys. Rev. D},
    volume        = {69},
    pages         = {044026},
    year          = {2004},
    doi           = {10.1103/PhysRevD.69.044026}
}

@article{Cooney:2009rr,
    author        = {Cooney, Alan and DeDeo, Simon and Psaltis, Dimitrios},
    title         = {Neutron Stars in $f(R)$ Gravity with Perturbative Constraints},
    eprint        = {0910.5480},
    archivePrefix = {arXiv},
    primaryClass  = {astro-ph.HE},
    journal       = {Phys. Rev. D},
    volume        = {82},
    pages         = {064033},
    year          = {2010},
    doi           = {10.1103/PhysRevD.82.064033}
}

@article{Arapoglu:2010rz,
    author        = {Arapoglu, A. S. and Deliduman, C. and Eksi, K. Y.},
    title         = {Constraints on Perturbative $f(R)$ Gravity via Neutron Stars},
    eprint        = {1003.3179},
    archivePrefix = {arXiv},
    primaryClass  = {gr-qc},
    journal       = {JCAP},
    volume        = {07},
    pages         = {020},
    year          = {2011},
    doi           = {10.1088/1475-7516/2011/07/020}
}

@article{Orellana:2013gn,
    author        = {Orellana, Mariana and Garcia, Federico and Teppa Pannia, Florencia A. and Romero, Gustavo E.},
    title         = {Structure of Neutron Stars in $R$-Squared Gravity},
    eprint        = {1301.5189},
    archivePrefix = {arXiv},
    primaryClass  = {astro-ph.CO},
    journal       = {Gen. Rel. Grav.},
    volume        = {45},
    pages         = {771--783},
    year          = {2013},
    doi           = {10.1007/s10714-013-1501-5}
}

@article{Damour:1992we,
    author  = {Damour, Thibault and Esposito-Farese, Gilles},
    title   = {Tensor multiscalar theories of gravitation},
    journal = {Class. Quant. Grav.},
    volume  = {9},
    pages   = {2093--2176},
    year    = {1992},
    doi     = {10.1088/0264-9381/9/9/015}
}

@article{Damour:1993hw,
    author  = {Damour, Thibault and Esposito-Farese, Gilles},
    title   = {Nonperturbative strong-field effects in tensor-scalar theories of gravitation},
    journal = {Phys. Rev. Lett.},
    volume  = {70},
    pages   = {2220--2223},
    year    = {1993},
    doi     = {10.1103/PhysRevLett.70.2220}
}

@article{Carroll:2003wy,
    author = {Carroll, Sean M. and Duvvuri, Vikram and Trodden, Mark and Turner, Michael S.},
    title = {Is Cosmic Speed-Up Due to New Gravitational Physics?},
    eprint = {astro-ph/0306438},
    archivePrefix = {arXiv},
    doi = {10.1103/PhysRevD.70.043528},
    journal = {Phys. Rev. D},
    volume = {70},
    pages = {043528},
    year = {2004}
}

@article{Hu:2007nk,
    author = {Hu, Wayne and Sawicki, Ignacy},
    title = {Models of $f(R)$ Cosmic Acceleration that Evade Solar-System Tests},
    eprint = {0705.1158},
    archivePrefix = {arXiv},
    primaryClass = {astro-ph},
    doi = {10.1103/PhysRevD.76.064004},
    journal = {Phys. Rev. D},
    volume = {76},
    pages = {064004},
    year = {2007}
}

@article{Starobinsky:1980te,
    author  = {Starobinsky, Alexei A.},
    title   = {A New Type of Isotropic Cosmological Models Without Singularity},
    journal = {Phys. Lett. B},
    volume  = {91},
    pages   = {99--102},
    year    = {1980},
    doi     = {10.1016/0370-2693(80)90670-X}
}

@article{Donoghue:1994dn,
    author        = {Donoghue, John F.},
    title         = {General Relativity as an Effective Field Theory:
                     The Leading Quantum Corrections},
    eprint        = {gr-qc/9405057},
    archivePrefix = {arXiv},
    journal       = {Phys. Rev. D},
    volume        = {50},
    pages         = {3874--3888},
    year          = {1994},
    doi           = {10.1103/PhysRevD.50.3874}
}

@article{Burgess:2003jk,
    author        = {Burgess, C. P.},
    title         = {Quantum Gravity in Everyday Life:
                     General Relativity as an Effective Field Theory},
    eprint        = {gr-qc/0311082},
    archivePrefix = {arXiv},
    journal       = {Living Rev. Rel.},
    volume        = {7},
    pages         = {5},
    year          = {2004},
    doi           = {10.12942/lrr-2004-5}
}

@misc{Gupta:2026mqk,
    author = "Gupta, Pawan Kumar",
    title = "{Binary Black Holes in Einstein-Maxwell-Dilaton Theory: Second Post-Newtonian Dynamics from Effective Field Theory}",
    eprint = "2609.07413",
    archivePrefix = "arXiv",
    primaryClass = "gr-qc",
    month = "9",
    year = "2026"
}

@article{Gupta:2022spq,
    author = "Gupta, Pawan Kumar",
    title = "{Binary dynamics from Einstein-Maxwell theory at second post-Newtonian order using effective field theory}",
    eprint = "2205.11591",
    archivePrefix = "arXiv",
    primaryClass = "gr-qc",
    doi = "10.1103/p1y1-1kcr",
    journal = "Phys. Rev. D",
    volume = "112",
    number = "10",
    pages = "104047",
    year = "2025"
}

@misc{LIGOScientific:2026fcf,
    author = "Abac, A. G. and others",
    collaboration = "LIGO Scientific, VIRGO, KAGRA",
    title = "{GWTC-4.0: Tests of General Relativity. II. Parameterized Tests}",
    eprint = "2603.19020",
    archivePrefix = "arXiv",
    primaryClass = "gr-qc",
    reportNumber = "LIGO-P2500066",
    month = "3",
    year = "2026"
}

@misc{LIGOScientific:2026gwtc5gr,
    author = "Abac, A. G. and others",
    collaboration = "LIGO Scientific, Virgo, KAGRA",
    title = "{GWTC-5.0: Tests of General Relativity}",
    eprint = "2607.19293",
    archivePrefix = "arXiv",
    primaryClass = "gr-qc",
    reportNumber = "LIGO-P2500781",
    year = "2026"
}

@article{vanGemeren:2023rhh,
    author = "van Gemeren, Iris and Shiralilou, Banafsheh and Hinderer, Tanja",
    title = "{Dipolar tidal effects in gravitational waves from scalarized black hole binary inspirals in quadratic gravity}",
    eprint = "2302.08480",
    archivePrefix = "arXiv",
    primaryClass = "gr-qc",
    doi = "10.1103/PhysRevD.108.024026",
    journal = "Phys. Rev. D",
    volume = "108",
    number = "2",
    pages = "024026",
    year = "2023",
    note = "[Erratum: Phys.Rev.D 109, 089901 (2024)]"
}

@article{KAGRA:2018plz,
    author = "Akutsu, T. and others",
    collaboration = "KAGRA",
    title = "{KAGRA: 2.5 Generation Interferometric Gravitational Wave Detector}",
    eprint = "1811.08079",
    archivePrefix = "arXiv",
    primaryClass = "gr-qc",
    doi = "10.1038/s41550-018-0658-y",
    journal = "Nature Astron.",
    volume = "3",
    number = "1",
    pages = "35--40",
    year = "2019"
}

@article{Berti:2018cxi,
    author = "Berti, Emanuele and others",
    title = "{Extreme gravity tests with gravitational waves from compact binary coalescences}",
    eprint = "1801.03587",
    archivePrefix = "arXiv",
    primaryClass = "gr-qc",
    doi = "10.1088/1361-6382/aac9f0",
    journal = "Class. Quant. Grav.",
    volume = "35",
    pages = "063001",
    year = "2018"
}

@article{Sotiriou:2015pka,
    author = "Sotiriou, Thomas P.",
    title = "{Black Holes and Scalar Fields}",
    eprint = "1505.00248",
    archivePrefix = "arXiv",
    primaryClass = "gr-qc",
    doi = "10.1088/0264-9381/32/21/214002",
    journal = "Class. Quant. Grav.",
    volume = "32",
    pages = "214002",
    year = "2015"
}

@article{Roy:2025xih,
    author = "Roy, Soumen and Janquart, Justin",
    title = "{Testing modified gravity with the eccentric neutron star-black hole merger GW200105}",
    eprint = "2507.21315",
    archivePrefix = "arXiv",
    primaryClass = "gr-qc",
    reportNumber = "LIGO-P2500438",
    doi = "10.1103/6scp-xqhb",
    journal = "Phys. Rev. D",
    volume = "113",
    number = "2",
    pages = "024056",
    year = "2026"
}

@article{Yagi:2011xp,
    author = "Yagi, Kent and Tanaka, Takahiro",
    title = "{Constraining alternative theories of gravity by gravitational waves from compact binaries in Einstein-Dilaton-Gauss-Bonnet gravity}",
    eprint = "1108.3283",
    archivePrefix = "arXiv",
    primaryClass = "gr-qc",
    doi = "10.1103/PhysRevD.86.044037",
    journal = "Phys. Rev. D",
    volume = "86",
    pages = "044037",
    year = "2012"
}

@article{Yagi:2011ze,
    author = "Yagi, Kent and Yunes, Nicolas and Tanaka, Takahiro",
    title = "{Gravitational Waves from Quasi-Circular Black Hole Binaries in Einstein-Dilaton-Gauss-Bonnet Gravity}",
    eprint = "1110.5950",
    archivePrefix = "arXiv",
    primaryClass = "gr-qc",
    doi = "10.1103/PhysRevLett.107.251103",
    journal = "Phys. Rev. Lett.",
    volume = "107",
    pages = "251103",
    year = "2011"
}

@article{Yagi:2012vf,
    author = "Yagi, Kent and Yunes, Nicolas and Tanaka, Takahiro",
    title = "{Gravitational Waves from Quasi-Circular Black Hole Binaries in Dynamical Chern-Simons Gravity}",
    eprint = "1208.5102",
    archivePrefix = "arXiv",
    primaryClass = "gr-qc",
    doi = "10.1103/PhysRevD.86.044037",
    journal = "Phys. Rev. D",
    volume = "86",
    pages = "044037",
    year = "2012"
}

@article{Yunes:2011we,
    author = "Yunes, Nicolas and Pani, Paolo and Cardoso, Vitor",
    title = "{Gravitational waves from quasicircular extreme mass-ratio inspirals as probes of scalar-tensor theories}",
    eprint = "1112.3351",
    archivePrefix = "arXiv",
    primaryClass = "gr-qc",
    doi = "10.1103/PhysRevD.85.102003",
    journal = "Phys. Rev. D",
    volume = "85",
    pages = "102003",
    year = "2012"
}

@article{Yagi:2012ya,
    author = "Yagi, Kent and Stein, Leo C. and Yunes, Nicolas and Tanaka, Takahiro",
    title = "{Post-Newtonian, Quasi-Circular Binary Inspirals in Quadratic Modified Gravity}",
    eprint = "1208.5102",
    archivePrefix = "arXiv",
    primaryClass = "gr-qc",
    doi = "10.1103/PhysRevD.85.064022",
    journal = "Phys. Rev. D",
    volume = "85",
    pages = "064022",
    year = "2012"
}

@article{Silva:2017uqg,
    author = "Silva, Hector O. and others",
    title = "{Spontaneous scalarization of black holes and compact stars from a Gauss-Bonnet coupling}",
    eprint = "1711.02080",
    archivePrefix = "arXiv",
    primaryClass = "gr-qc",
    doi = "10.1103/PhysRevLett.120.131104",
    journal = "Phys. Rev. Lett.",
    volume = "120",
    pages = "131104",
    year = "2018"
}

@article{LIGOScientific:2019fpa,
    author = "Abbott, B. P. and others",
    collaboration = "LIGO Scientific and Virgo",
    title = "{Tests of General Relativity with Binary Black Hole Signals from the LIGO-Virgo Catalog GWTC-1}",
    eprint = "1903.04467",
    archivePrefix = "arXiv",
    primaryClass = "gr-qc",
    doi = "10.1103/PhysRevD.100.104036",
    journal = "Phys. Rev. D",
    volume = "100",
    pages = "104036",
    year = "2019"
}

@article{LIGOScientific:2021sio,
    author = "Abbott, R. and others",
    collaboration = "LIGO Scientific and Virgo",
    title = "{Tests of General Relativity with GWTC-2}",
    eprint = "2010.14529",
    archivePrefix = "arXiv",
    primaryClass = "gr-qc",
    doi = "10.1103/PhysRevD.103.122002",
    journal = "Phys. Rev. D",
    volume = "103",
    pages = "122002",
    year = "2021"
}

@article{LIGOScientific:2021nrg,
    author = "Abbott, R. and others",
    collaboration = "LIGO Scientific, Virgo, and KAGRA",
    title = "{Tests of General Relativity with GWTC-3}",
    eprint = "2112.06861",
    archivePrefix = "arXiv",
    primaryClass = "gr-qc",
    doi = "10.1103/PhysRevD.106.044028",
    journal = "Phys. Rev. D",
    volume = "106",
    pages = "044028",
    year = "2022"
}

@article{Yunes:2009ke,
    author = "Yunes, Nicolas and Pretorius, Frans",
    title = "{Fundamental Theoretical Bias in Gravitational Wave Astrophysics and the Parameterized Post-Einsteinian Framework}",
    eprint = "0909.3328",
    archivePrefix = "arXiv",
    primaryClass = "gr-qc",
    doi = "10.1103/PhysRevD.80.122003",
    journal = "Phys. Rev. D",
    volume = "80",
    pages = "122003",
    year = "2009"
}

@article{Mishra:2010tp,
    author = "Mishra, Chandra Kant and Arun, K. G. and Iyer, Bala R. and Sathyaprakash, B. S.",
    title = "{Parametrized tests of post-Newtonian theory using Advanced LIGO and Einstein Telescope}",
    eprint = "1005.0304",
    archivePrefix = "arXiv",
    primaryClass = "gr-qc",
    doi = "10.1103/PhysRevD.82.064010",
    journal = "Phys. Rev. D",
    volume = "82",
    pages = "064010",
    year = "2010"
}

@article{Li:2011cg,
    author = "Li, T. G. F. and Del Pozzo, W. and Vitale, S. and Van Den Broeck, C. and Agathos, M. and Veitch, J. and Grover, K. and Sidery, T. and Sturani, R. and Vecchio, A.",
    title = "{Towards a generic test of the strong field dynamics of general relativity using compact binary coalescence}",
    eprint = "1110.0530",
    archivePrefix = "arXiv",
    primaryClass = "gr-qc",
    doi = "10.1103/PhysRevD.85.082003",
    journal = "Phys. Rev. D",
    volume = "85",
    pages = "082003",
    year = "2012"
}

@article{Arun:2006hn,
    author = "Arun, K. G. and Iyer, Bala R. and Qusailah, M. S. S. and Sathyaprakash, B. S.",
    title = "{Probing the non-linear structure of general relativity with black hole binaries}",
    eprint = "gr-qc/0604067",
    archivePrefix = "arXiv",
    doi = "10.1103/PhysRevD.74.024006",
    journal = "Phys. Rev. D",
    volume = "74",
    pages = "024006",
    year = "2006"
}

@article{Shoom:2021mdj,
    author = "Shoom, Andrey A. and Gupta, Pawan K. and Krishnan, Badri and Nielsen, Alex B. and Capano, Collin D.",
    title = "{Testing the post-Newtonian expansion with GW170817}",
    eprint = "2105.02191",
    archivePrefix = "arXiv",
    primaryClass = "gr-qc",
    doi = "10.1007/s10714-023-03100-z",
    journal = "Gen. Rel. Grav.",
    volume = "55",
    number = "4",
    pages = "55",
    year = "2023"
}

@article{Will:2014kxa,
    author = "Will, Clifford M.",
    title = "{The Confrontation between General Relativity and Experiment}",
    eprint = "1403.7377",
    archivePrefix = "arXiv",
    primaryClass = "gr-qc",
    doi = "10.12942/lrr-2014-4",
    journal = "Living Rev. Rel.",
    volume = "17",
    pages = "4",
    year = "2014"
}

@article{Berti:2015itd,
    author = "Berti, Emanuele and others",
    title = "{Testing General Relativity with Present and Future Astrophysical Observations}",
    eprint = "1501.07274",
    archivePrefix = "arXiv",
    primaryClass = "gr-qc",
    doi = "10.1088/0264-9381/32/24/243001",
    journal = "Class. Quant. Grav.",
    volume = "32",
    pages = "243001",
    year = "2015"
}

@article{Yunes:2016jcc,
    author = "Yunes, Nicolás and Yagi, Kent and Pretorius, Frans",
    title = "{Theoretical Physics Implications of the Binary Black-Hole Mergers GW150914 and GW151226}",
    eprint = "1603.08955",
    archivePrefix = "arXiv",
    primaryClass = "gr-qc",
    doi = "10.1103/PhysRevD.94.084002",
    journal = "Phys. Rev. D",
    volume = "94",
    number = "8",
    pages = "084002",
    year = "2016"
}

@article{LIGOScientific:2020tif,
    author = "Abbott, R. and others",
    collaboration = "LIGO Scientific and Virgo",
    title = "{GWTC-2: Compact Binary Coalescences Observed by LIGO and Virgo During the First Half of the Third Observing Run}",
    eprint = "2010.14527",
    archivePrefix = "arXiv",
    primaryClass = "gr-qc",
    doi = "10.1103/PhysRevX.11.021053",
    journal = "Phys. Rev. X",
    volume = "11",
    number = "2",
    pages = "021053",
    year = "2021"
}

@article{LIGOScientific:2016lio,
    author = "Abbott, B. P. and others",
    collaboration = "LIGO Scientific and Virgo",
    title = "{GW150914: First results from the search for binary black hole coalescence with Advanced LIGO}",
    eprint = "1602.03839",
    archivePrefix = "arXiv",
    primaryClass = "gr-qc",
    doi = "10.1103/PhysRevD.93.122003",
    journal = "Phys. Rev. D",
    volume = "93",
    number = "12",
    pages = "122003",
    year = "2016"
}

@article{Gupta:2021rod,
    author = "Gupta, Pawan Kumar and Spieksma, Thomas F. M. and Pang, Peter T. H. and Koekoek, Gideon and Broeck, Chris Van Den",
    title = "{Bounding dark charges on binary black holes using gravitational waves}",
    eprint = "2107.12111",
    archivePrefix = "arXiv",
    primaryClass = "gr-qc",
    doi = "10.1103/PhysRevD.104.063041",
    journal = "Phys. Rev. D",
    volume = "104",
    number = "6",
    pages = "063041",
    year = "2021"
}

@article{LIGOScientific:2014pky,
    author = "Aasi, J. and others",
    collaboration = "LIGO Scientific",
    title = "{Advanced LIGO}",
    eprint = "1411.4547",
    archivePrefix = "arXiv",
    primaryClass = "gr-qc",
    doi = "10.1088/0264-9381/32/7/074001",
    journal = "Class. Quant. Grav.",
    volume = "32",
    pages = "074001",
    year = "2015"
}

@article{VIRGO:2014yos,
    author = "Acernese, F. and others",
    collaboration = "VIRGO",
    title = "{Advanced Virgo: a second-generation interferometric gravitational wave detector}",
    eprint = "1408.3978",
    archivePrefix = "arXiv",
    primaryClass = "gr-qc",
    doi = "10.1088/0264-9381/32/2/024001",
    journal = "Class. Quant. Grav.",
    volume = "32",
    number = "2",
    pages = "024001",
    year = "2015"
}

\clearpage
\onecolumngrid

\appendix

\setcounter{secnumdepth}{3}
\setcounter{section}{0}
\renewcommand{\thesection}{S\arabic{section}}

\begin{center}
{\Large\bfseries SUPPLEMENTAL MATERIAL}
\end{center}

\setcounter{equation}{0}

\renewcommand{\theequation}{S\arabic{equation}}

\section{Frequency-domain phase in $R^2$ gravity}
\label{app:R2_phase}
We construct the frequency-domain inspiral waveform in quadratic
$f(R)=R+aR^2$ gravity by mapping the theory onto the EFT description of generic massive scalar-tensor gravity developed
in Ref.~\cite{Diedrichs:2023}. The GW phase can be
expressed as
\begin{equation}
    \Psi(f)
    =
    \Psi_{\rm GR}(f)
    +
    \Psi_{\rm ST}(f),
    \label{eq:app_total_phase}
\end{equation}
where $\Psi_{\rm GR}$ denotes the GR phase, and $\Psi_{\rm ST}$ represents
the additional contributions arising from the massive scalar field. These contributions can be decomposed into conservative and radiative
contributions,
\begin{equation}
\begin{split}
    \Psi_{\rm ST}(f)
    ={}&
    \Psi_E(f)
    +
    \Theta\left(
        \frac{v^3}{\bar \mu_s}-1
    \right)\Psi_{\ell=1}(f) +
    \Theta\left(
        \frac{v^3}{\bar \mu_s}-\frac{1}{2}
    \right)\Psi_{\ell=2}(f),
\end{split}
\label{eq:app_phase_decomposition}
\end{equation}
where $\Theta$ denotes the Heaviside step function and
\begin{equation}
    v=
    \left(\frac{\pi GMf}{c^3}\right)^{1/3},
    \qquad
    \bar \mu_s=\frac{GMm_s}{\hbar c},
    \qquad
    M=m_1+m_2.
    \label{eq:app_dimensionless_mass}
\end{equation}
Here $m_s$ denotes the scalar mass expressed in energy units
divided by $c^2$.

The conservative contribution $\Psi_E$ arises from the
scalar-mediated modification of the binary interaction and
enters at leading order at $0$PN. In contrast, the leading
scalar-dipole radiation contributes at $-1$PN, while the
scalar-quadrupole radiation enters at $0$PN. Unlike the conservative correction, scalar radiation is
subject to a kinematic threshold determined by the scalar
mass. The dipole and quadrupole channels become accessible
at frequencies
\begin{equation}
    f_{\ell=1}
    =
    \frac{m_sc^2}{\pi\hbar},
    \qquad
    f_{\ell=2}
    =
    \frac{m_sc^2}{2\pi\hbar},
    \label{eq:app_activation_frequency}
\end{equation}
respectively. Below, we give explicit expressions for the leading conservative, dipole, and quadrupole phase corrections and define the corresponding EFT parameters. We use natural units, $c=\hbar=1$, in these expressions and in Table~\ref{tab:dimenionless_params}. 

\begin{subequations} 
    \begin{align}
    \begin{split}
    \label{eq:phase_LO}
        \Psi_E^{\text{0PN}} &= \frac{5 \bar{Q}}{6 \nu  \bar{\mu}_s^{5/2}} \left[\Gamma \left(\frac{5}{2},\frac{\bar{\mu}_s}{v^2}\right)+\Gamma \left(\frac{7}{2},\frac{\bar{\mu}_s}{v^2}\right)+2 \, \Gamma
        \left(\frac{9}{2},\frac{\bar{\mu}_s}{v^2}\right)\right] \\ 
        &\hspace{5cm} - 
         v^3 \frac{5 \bar{Q}}{6 \nu
         \bar{\mu}_s^4} \bigg[\Gamma \bigg(4,\frac{\bar{\mu}_s}{v^2}\bigg)+\Gamma
         \bigg(5,\frac{\bar{\mu}_s}{v^2}\bigg) + 2\,\Gamma \bigg(6,\frac{\bar{\mu}_s}{v^2}\bigg)\bigg] \; , 
    \end{split}\\[0.6cm]
    \label{eq:phase_LO1}
        \Psi_{\ell=1}^{-1\text{PN}} &= \delta  q^2 \frac{5}{896 \nu ^3} \left[-\frac{1}{v^7} \,{}_3F_2\left(-\frac{3}{2},\frac{5}{3},\frac{7}{6} ; \frac{8}{3},\frac{13}{6};\frac{\bar{\mu}_s^2}{v^6}\right) - \frac{90 \cdot 2^{1/3} 3^{1/2} v^3 \Gamma \left(\frac{2}{3}\right)^3 }{247 \pi \bar{\mu}_s^{10/3}} + \frac{63 \, \Gamma \left(\frac{1}{3}\right)^3}{256 \cdot 2^{1/3} \pi \bar{\mu}_s^{7/3}} \right] \; .
    \end{align}
\end{subequations}
Here we utilized the \textit{upper} incomplete gamma function $\Gamma(a, x) = \int_x^\infty t^{a-1} e^{-t} dt$ and ${}_3F_2$ denotes a generalized hypergeometric function.
Note that the above expressions only contain modifications due to the scalar charges $q_1$ and $q_2$.
Corrections due to the induced charges $p_1$ and $p_2$, as well as the self-interaction parameter $c_3$, arise at next-to-leading order (NLO, $0$PN) with respect to the scalar radiation. Explicitly, these terms are given by
\begin{subequations} \label{eq:phase_NLO}
    \begin{align}
        \begin{split}
            \Psi_{\ell=1}^{\text{0PN},q} &= -\frac{5 g_2 \delta q \bar{\mu}_s^2}{9856 \nu^3 v^{11}} \,{}_3F_2\left(-\frac{3}{2},\frac{7}{3},\frac{11}{6} ; \frac{10}{3},\frac{17}{6} ; \frac{\bar{\mu}_s^2}{v^6}\right) + \frac{45 \Gamma\left(\frac{2}{3}\right)^3 \delta q \left(-1680 g_1- 966 g_2+5 (1064 \nu +659) \delta q\right)}{1605632\ 2^{2/3} \pi  \nu ^3 \bar{\mu}_s^{5/3}} \\
            &{}\hspace{0.4cm} + \frac{\delta q \left(1680 g_1 + 1008 g_2 - 5 (1064 \nu +659) \delta q\right)}{86016 \nu ^3 v^5} \,{}_3F_2 \left(-\frac{3}{2},\frac{4}{3},\frac{5}{6} ; \frac{7}{3},\frac{11}{6} ; \frac{\bar{\mu}_s^2}{v^6}\right) \\
            &{}\hspace{0.4cm}+ \frac{5 \sqrt{3} v^3 \Gamma \left(\frac{1}{3}\right)^3 \delta q \left(7728 g_1+ 4368 g_2-23 (1064 \nu + 659) \delta q\right)}{30829568 \sqrt[3]{2} \pi  \nu ^3 \bar{\mu}_s^{8/3}} \; ,
        \end{split}\\[0.6cm]
        \begin{split}
            \Psi_{\ell=1}^{\text{0PN},p} &= \Xi_p \frac{5}{16 \nu^3} \left[- \frac{1}{v^5} \,{}_3F_2\left(-\frac{3}{2},\frac{4}{3},\frac{5}{6} ; \frac{7}{3},\frac{11}{6};\frac{\bar{\mu}_s^2}{v^6}\right) - \frac{6 \cdot 2^{2/3} 3^{1/2} v^3 \Gamma \left(\frac{1}{3}\right)^3 }{187 \pi \bar{\mu}_s^{8/3}} + \frac{135 \, \Gamma \left(\frac{2}{3}\right)^3}{56 \cdot 2^{2/3} \pi \bar{\mu}_s^{5/3}} \right] \; ,
        \end{split}\\[0.6cm]
        \begin{split} \label{c3::waveform}
            \Psi_{\ell=1}^{\text{0PN},c_3} &= \Xi_c \frac{5 G m_1 m_2}{7168 \pi \msbar \nu^3} \left[\frac{1}{v^7} \,{}_3F_2\left(-\frac{3}{2},\frac{5}{3},\frac{7}{6} ; \frac{8}{3},\frac{13}{6};\frac{\bar{\mu}_s^2}{v^6}\right) + \frac{90 \cdot 2^{1/3} 3^{1/2} v^3 \Gamma \left(\frac{2}{3}\right)^3 }{247 \pi \bar{\mu}_s^{10/3}} - \frac{63 \, \Gamma \left(\frac{1}{3}\right)^3}{256 \cdot 2^{1/3} \pi \bar{\mu}_s^{7/3}} \right] \\
            &{}\hspace{0.4cm}+ \Xi_c \frac{25 G m_1 m_2}{55296 \pi \nu^3} \left[ - \frac{1}{v^9} \,{}_3F_2\left(-\frac{3}{2}, 2,\frac{3}{2} ; 3,\frac{5}{2};\frac{\bar{\mu}_s^2}{v^6}\right) - \frac{24 v^3}{35 \bar{\mu}_s^4} + \frac{3 \pi}{8 \bar{\mu}_s^3} \right] \; ,
        \end{split}\\[0.6cm]
        \begin{split}
        \label{eq:quadpolar_term}
            \Psi_{\ell=2}^\text{0PN} &= \Xi_q^2 \frac{1}{32 \nu ^3} \left[-\frac{1}{v^5} \,{}_3F_2\left(-\frac{5}{2},\frac{4}{3},\frac{5}{6} ; \frac{7}{3},\frac{11}{6};\frac{\bar{\mu}_s^2}{4 v^6}\right) - \frac{720 \cdot 2^{1/3} 3^{1/2} v^3 \Gamma \left(\frac{1}{3}\right)^3 }{4301 \pi \bar{\mu}_s^{8/3}} + \frac{405 \, \Gamma \left(\frac{2}{3}\right)^3}{112 \pi \bar{\mu}_s^{5/3}} \right] \; .
        \end{split}
    \end{align}
\end{subequations}

Above, we denote contributions arising from a certain scalar field parameter with a respective superscript. For example, $\Psi_{\ell=1}^{\text{0PN},c_3}$ denotes the contribution of $c_3$ to the $0$PN order phase shift in dipole radiation. For the definition of all rescaled parameters, we refer to Table~\ref{tab:dimenionless_params}.

\begin{table*}[t]
  \centering
  \begin{tabular}{|c|c|c|c|c|c|c|c|c|c|c|}
  \hline
    $M$ & $x$ & $\gamma$ & $\mu$ & $\nu$ & $Q$ & $\bar{Q}$ & $\displaystyle \bar{\mu}_s$ & $\bar{c}_3$ &$\xi_q$ & $\xi_p$ 
    \\ \hline
    \rule{0pt}{0.65cm}
    $\displaystyle m_1 + m_2 $ & $\displaystyle (G M \omega)^\frac{2}{3}$ & $ \displaystyle \frac{G M}{r}$ & $\displaystyle \frac{m_1 m_2}{M}$ & $\displaystyle \frac{\mu}{M}$ &  $\displaystyle \frac{q_1+q_2}{M}$ & $\displaystyle \frac{q_1 q_2}{m_1 m_2}$ & $\displaystyle G M m_s$ & $\displaystyle c_3  G M^2  Q  \bar{Q}$& $ \displaystyle \frac{q_1^2 m_2 + q_2^2 m_1}{M m_1 m_2}$ & $\displaystyle \frac{p_1 q_2^2 + p_2 q_1^2}{M m_1 m_2}$ \rule[-0.5cm]{0pt}{0pt} \\ 
    \hline 
  \end{tabular}
  \newline
  \vspace{0.01cm}
  \newline
  \setlength\tabcolsep{4.5pt}
  \begin{tabular}{|c|c|c|c|c|c|c|c|}
  \hline \rule{0pt}{0.45cm}
    $v$ & $\delta q$ & $\delta M$ & $\Xi_p$ & $\Xi_c$ & $\Xi_q$ & $g_1$ & $g_2$ \\ \hline
    \rule{0pt}{0.65cm}
    $(\pi G M f)^\frac{1}{3}$&$\displaystyle \frac{m_2 q_1-m_1 q_2}{M^2}$ & $ \displaystyle m_1 - m_2$ & $\displaystyle \frac{m_1 p_2 q_1 - m_2 p_1 q_2}{M^3} \delta q$ & $\displaystyle \frac{c_3 \bar{Q} \delta M \delta q}{M}$ &  $\displaystyle \frac{m_2^2 q_1 + m_1^2 q_2}{M^3}$ & 
    $\displaystyle \frac{m_2^2 q_1 - m_1^2 q_2}{M^3}$ 
    & $\displaystyle \frac{m_2^3 q_1 - m_1^3 q_2}{M^4}$  \rule[-0.5cm]{0pt}{0pt}\\ \hline 
  \end{tabular}
  \caption{Overview of the parameters introduced in our work. Besides the typical PN expansion variables, we rescale the EFT parameters for convenience.}
  \label{tab:dimenionless_params}
\end{table*}

\subsection{Binary neutron-star systems}
\label{app:BNS_phase}

For BNS systems, the scalar charges, induced charges, and scalar mass in quadratic
$f(R)$ gravity are given by Eqs.~\eqref{eq:worldline_coefficients} and~\eqref{eq:self_couplings}
\begin{equation}
    q_A=-\frac{m_A}{2\sqrt{6}},
    \qquad
    p_A= \frac{m_A}{48},
    \qquad
    m_s=\frac{1}{\sqrt{6a}},
    \label{eq:app_R2_matching_BNS}
\end{equation}
which yield the corresponding dimensionless scalar coupling and the dipole-charge parameter,
\begin{equation}
    \bar Q
    =
    \frac{q_1q_2}{m_1m_2}
    =
    \frac{1}{24}, \quad \delta q
    =
    \frac{m_2q_1-m_1q_2}{M^2}
    =
    0.
    \label{eq:app_Q_R2_BNS}
\end{equation}
Since the scalar-dipole phase corrections are proportional
to $\delta q$, the leading and next-to-leading dipole
contributions vanish. In particular,
\begin{equation}
    \Psi_{\ell=1}^{-1\mathrm{PN}}
    =
    \Psi_{\ell=1}^{0\mathrm{PN}}
    =
    0.
\end{equation}
Thus, in $R^2$ gravity the $-1$PN and $0$PN scalar-dipole corrections are absent in BNS systems.

\begin{figure}[t]
    \centering
    \includegraphics[width=0.5\columnwidth]
    {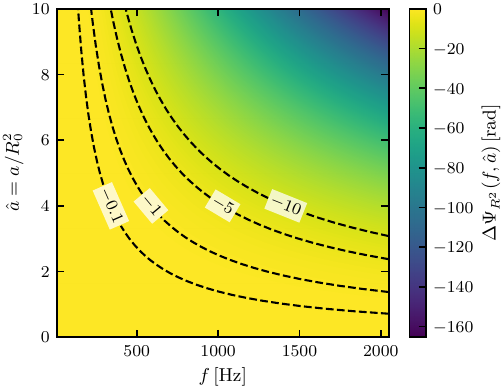}
    \caption{
    Phase correction $\Delta\Psi_{R^2}$ in quadratic
    $f(R)$ gravity for a $(1.4,\,1.3)\,M_\odot$ BNS, as a function of GW
    frequency $f$ and coupling $\hat a$.
    Dashed curves are contours of constant $\Delta\Psi_{R^2}$ (in rad).
    }
    \label{fig:phase_2D_map}
\end{figure}

The remaining scalar contributions arise from the conservative
interaction and scalar-quadrupole radiation, giving
\begin{equation}
    \begin{split}
    \Psi_{R^2}^{\rm BNS}(f) &= \Psi_{\rm GR}(f) + \Delta\Psi_{R^2}(f;a), \\
    \Delta\Psi_{R^2}(f;a) &= \Psi_{R^2,E}^{0\mathrm{PN}}(f)
        + \Theta\left(
        \frac{v^3}{\bar{\mu}_s}-\frac{1}{2}
    \right)
    \Psi_{R^2,\ell=2}^{0\mathrm{PN}}(f)
    \end{split}
    \label{eq:app_complete_R2_phase_BNS}
\end{equation}
The conservative contribution
$\Psi_{R^2,E}^{0\mathrm{PN}}$
is obtained from Eq.~\eqref{eq:phase_LO}
by substituting $\bar Q=1/24$.
Similarly, the scalar-quadrupole contribution follows
from Eq.~\eqref{eq:quadpolar_term} with
\begin{equation}
    \Xi_q
    =
    -\frac{1}{2\sqrt{6}}
    \frac{m_1m_2}{M^2}
    =
    -\frac{\nu}{2\sqrt{6}}.
    \label{eq:app_Xiq_BNS}
\end{equation}
Scalar-quadrupole radiation requires $f \geq f_{\ell=2}$ (Eq.~\eqref{eq:app_activation_frequency}), whereas the
conservative correction is not
subject to any frequency threshold. The phase correction $\Delta\Psi_{R^2}(f;a)$ depends strongly on the coupling $\hat a$ (Fig.~\ref{fig:phase_2D_map}); as $\hat a$ changes from 0 to 10, its magnitude can reach $\sim 100$ radians at higher frequency, which gives a hint about our prior range of $\hat a$ for the Bayesian analysis.

\subsection{Neutron-star--black-hole systems}
\label{app:NSBH_phase}

For NSBH binaries, the scalar-charge configuration
differs qualitatively from that of BNS systems. As a black hole carries no scalar charge in $R^2$ gravity, the scalar charges for NSBH given by Eq.~\eqref{eq:worldline_coefficients},
\begin{equation}
    q_1=0,
    \qquad
    q_2=-\frac{m_2}{2\sqrt{6}},
    \qquad
    p_1= 0,
    \qquad
    p_2= \frac{m_2}{48}.
    \label{eq:app_R2_matching_NSBH}
\end{equation}
The vanishing black-hole scalar charge implies
\begin{equation}
    \bar Q
    =
    \frac{q_1q_2}{m_1m_2}
    =
    0.
    \label{eq:app_Q_R2_NSBH}
\end{equation}
Consequently, the leading conservative scalar
correction proportional to $\bar Q$ vanishes,
\begin{equation}
    \Psi_{R^2,E}^{0\mathrm{PN}}=0.
\end{equation}
In contrast to the BNS case, the unequal
scalar charge-to-mass ratios produce a
nonvanishing dipole-charge parameter,
\begin{equation}
    \delta q
    =
    \frac{m_2q_1-m_1q_2}{M^2}
    =
    \frac{\nu}{2\sqrt{6}}.
    \label{eq:app_delta_q_NSBH}
\end{equation}
Therefore, scalar-dipole radiation contributes
at leading $-1$PN order, together with
the next-to-leading $0$PN correction.

The corresponding frequency-domain phase is
\begin{equation}
\begin{split}
    \Psi_{R^2}^{\rm NSBH}(f)
    ={}&
    \Psi_{\rm GR}(f)
+\Delta\Psi_{R^2}(f; a)\\
    ={}&
    \Psi_{\rm GR}(f)
    +
    \Theta\left(
        \frac{v^3}{\bar \mu_s}-1
    \right)
    \left[
        \Psi_{R^2,\ell=1}^{-1\mathrm{PN}}(f)
        +
        \Psi_{R^2,\ell=1}^{0\mathrm{PN},q}(f)
    \right]
    +
    \Theta\left(
        \frac{v^3}{\bar \mu_s}-\frac{1}{2}
    \right)
    \Psi_{R^2,\ell=2}^{0\mathrm{PN}}(f).
\end{split}
\label{eq:app_complete_R2_phase_NSBH}
\end{equation}
Here the dipole contributions are obtained
from Eqs.~\eqref{eq:phase_LO1} and
\eqref{eq:phase_NLO} using the corresponding
NSBH scalar charges.

The scalar-quadrupole correction follows
from Eq.~\eqref{eq:quadpolar_term} with
\begin{equation}
    \Xi_q
    =
    -\frac{1}{2\sqrt{6}}
    \frac{m_1^2m_2}{M^3}.
    \label{eq:app_Xiq_NSBH}
\end{equation}
Both radiative contributions are subject
to their respective scalar-mass thresholds.
In particular, dipole radiation becomes
kinematically accessible only when
$f\geq f_{\ell=1}$ (Eq.~\eqref{eq:app_activation_frequency}), whereas quadrupole
radiation can occur for $f\geq f_{\ell=2}$ (Eq.~\eqref{eq:app_activation_frequency}). The phase correction $\Delta\Psi_{R^2}(f;a)$ depends weakly on the coupling $\hat a$ (Fig.~\ref{fig:phase_2D_map}); as $\hat a$ changes from 0 to 3000, it can increase up to a few radians at higher frequency.

\begin{figure}[t]
    \centering
    \includegraphics[width=0.5\columnwidth]
    {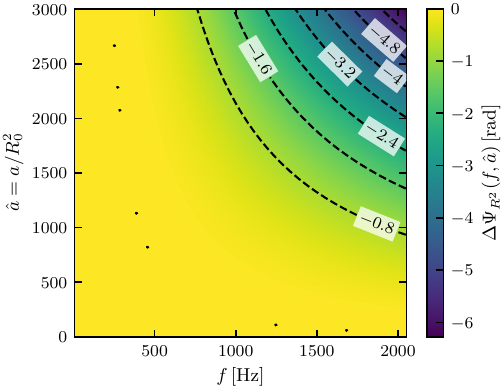}
    \caption{
    Phase correction $\Delta\Psi_{R^2}$ in quadratic
    $f(R)$ gravity for a $(8,\,1.4)\,M_\odot$ NSBH, as a function of GW
    frequency $f$ and coupling $\hat a$.
    Dashed curves are contours of constant $\Delta\Psi_{R^2}$ (in rad).
    }
    \label{fig:phase_2D_map_NSBH}
\end{figure}

This leads to an important distinction
between BNS and NSBH systems in the adopted
waveform model. For BNS binaries, the
conservative scalar correction can modify
the inspiral phase even below the scalar-radiation
thresholds. For NSBH binaries, the leading
conservative correction vanishes, leaving
the scalar-radiation channels as the dominant
effects retained at this order.

As illustrated in Fig.~\ref{fig:scalar_activation},
the activation frequencies depend on the
scalar mass and, consequently, on the coupling $a$. For sufficiently small
values of $a$, the scalar becomes heavy and
the dipole radiation thresholds $f_{\ell=1}$ and quadrupole radiation thresholds $f_{\ell=2}$ move beyond the
merger frequency of NSBH GW events. Therefore, within the approximations adopted
here, BNS inspirals can probe scalar-induced
conservative modifications in regions of
parameter space where the corresponding
NSBH radiative corrections are strongly
suppressed or kinematically inaccessible.



\end{document}